\documentclass[12pt]{article}

\pdfoutput=1
\usepackage{color}
\usepackage{epsfig, palatino}
\usepackage{epic}
\usepackage{dsfont}
\usepackage{colonequals}
\usepackage{mathrsfs}
\usepackage{ae} 
\usepackage[T1]{fontenc}
\usepackage[ansinew]{inputenc}
\usepackage{amsmath}
\usepackage{amssymb}
\usepackage{graphicx}
\usepackage[normalem]{ulem}
\usepackage{xcolor}
\definecolor{darkblue}{cmyk}{0.9,0.9,0,0}
\usepackage[colorlinks=true,linkcolor=darkblue,citecolor=darkblue,urlcolor=darkblue]{hyperref}
\usepackage{cite}
\usepackage{hyperref}
\usepackage{varioref}
\usepackage{makeidx}
\usepackage[english]{babel}
\usepackage{simplewick}
\usepackage{array}
\usepackage{multirow}

\usepackage[font={small}]{caption}

\newcommand{\comment}[1]{}

\newcommand{\begBvR}[1]{\begin{#1}} 
\newcommand{\beq}{\begBvR{equation}}
\newcommand{\eeq}{\end{equation}}

\newcommand{\eeqq}{\end{equation*}}

\newcommand\eeqaa{\end{eqnarray*}}

\newcommand\eeqa{\end{array}}
\newcommand{\bea}{\begBvR{eqnarray}}
\newcommand{\eea}{\end{eqnarray}}

\newcommand{\ud}{\mathrm d}

\newcommand{\neqa}{\nonumber\end{eqnarray}} 
\newcommand{\la}[1]{\label{#1}}

\newcommand{\ur}[1]{(\ref{#1})}

\renewcommand{\d}{\partial}

\newcommand{\<}{{\langle}}
\renewcommand{\>}{{\rangle}}

\newcommand{\re}{\relax{\rm I\kern-.18em R}}

\renewcommand{\sp}{p\hspace{-.40em}/}

\definecolor{darkgreen}{rgb}{0.0, 0.45, 0.0}
\definecolor{mathematicablue}{RGB}{94,130,182}

\def\XXint#1#2#3{{\setbox0=\hbox{$#1{#2#3}{\int}$}
\vcenter{\hbox{$#2#3$}}\kern-.5\wd0}}

\def\su2{{SU(2)}}

\def\a{{\alpha}}

\def\[{\left[}
\def\]{\right]}

\def\e{\epsilon}

\def\a{\alpha}

\def\({\left(}
\def\){\right)}
\def\[{\left[}
\def\]{\right]}

\def\<{\langle}
\def\>{\rangle}

\def\i2{\frac{i}{2}}

\def\spi{\relax{\rm \pi\kern-0.5em /}}
\def\sA{\relax{\rm A\kern-0.5em /}}
\def\sp{\relax{\rm p\kern-0.5em /}}
\def\sd{\relax{\rm \d\kern-0.5em /}}
\def\sk{\relax{\rm k\kern-0.5em /}}
\def\sn{\relax{\rm n\kern-0.5em /}}
\def\sl{\relax{\rm l\kern-0.5em /}}
\def\sP{\relax{\rm P\kern-0.7em /}}
\def\sBethe{\relax{\rm \Bethe\kern-0.5em /}}

\def\bp{{\bf p}}

\def\2F1{\,_2{\rm F}_1}

\makeindex

\newcommand\blfootnote[1]{%
  \begingroup
  \renewcommand\thefootnote{}\footnote{\hspace{-6mm}#1}%
  \addtocounter{footnote}{-1}%
  \endgroup
}

\begin{document}

\thispagestyle{empty}

\renewcommand{\thefootnote}{\fnsymbol{footnote}}
\setcounter{page}{1}
\setcounter{footnote}{0}
\setcounter{figure}{0}

\begin{flushright}
CERN-TH-2017-162
\end{flushright}
\vspace{-0.4in}
\begin{center}
$$$$
{\Large\textbf{\mathversion{bold}
The S-matrix Bootstrap III: \\ Higher Dimensional Amplitudes
}\par}
\vspace{1.0cm}

\textrm{Miguel F. Paulos$^\text{\tiny 1}$, Joao Penedones$^\text{\tiny 2}$, Jonathan Toledo$^\text{\tiny 2}$, Balt C. van Rees$^\text{\tiny 3}$, Pedro Vieira$^\text{\tiny 4,\tiny 5}$}
\blfootnote{\tt  \#@gmail.com\&/@\{miguel.paulos,jpenedones,jonathan.campbell.toledo,baltvanrees,pedrogvieira\}}
\\ \vspace{1.2cm}
\footnotesize{\textit{
$^\text{\tiny 1}$Theoretical Physics Department, CERN, Geneva, Switzerland\\
$^\text{\tiny 2}$
Institute of Physics, \'Ecole Polytechnique F\'ed\'erale de Lausanne (EPFL),
 CH-1015 Lausanne,
Switzerland\\
$^\text{\tiny 3}$Centre for Particle Theory, Department of Mathematical Sciences, Durham University, Lower Mountjoy, Stockton Road, Durham, England, DH1 3LE\\
$^\text{\tiny 4}$Perimeter Institute for Theoretical Physics,
Waterloo, Ontario N2L 2Y5, Canada\\
$^\text{\tiny 5}$ICTP South American Institute for Fundamental Research, IFT-UNESP, S\~ao Paulo, SP Brazil 01440-070  }  
\vspace{4mm}
}

\par\vspace{1.5cm}

\textbf{Abstract}\vspace{2mm}
\end{center}
We consider constraints on the S-matrix of any gapped, Lorentz invariant quantum field theory in 3+1 dimensions due to crossing symmetry, analyticity and unitarity. We extremize cubic couplings, quartic couplings and scattering lengths relevant for the elastic scattering amplitude of two identical scalar particles. In the cases where our results can be compared with the older S-matrix literature they are in excellent agreement. We also extremize a cubic coupling in 2+1 dimensions which we can directly compare to a universal bound for a QFT in AdS. This paper generalizes our previous 1+1 dimensional results of \cite{paperI} and \cite{paperII}.
\noindent

\setcounter{page}{1}
\renewcommand{\thefootnote}{\arabic{footnote}}
\setcounter{footnote}{0}

\setcounter{tocdepth}{2}

 \def\nref#1{{(\ref{#1})}}

\newpage

\tableofcontents

\parskip 5pt plus 1pt   \jot = 1.5ex

\newpage
\section{Introduction} 
In \cite{paperI} and \cite{paperII} we initiated a bootstrap analysis of massive quantum field theories. In particular, we obtained bounds on couplings of a  quantum field theory compatible with a given spectrum of stable particles. 

Physically, one expects such bounds to exist since increasing the interaction strength will typically increase the attraction between particles. As such, we expect to have maximum values for couplings beyond which the masses of bound states must decrease, or new bound-states should emerge from the continuum, or both.  

Mathematically, this problem is also very natural once we make the non-trivial assumption that scattering amplitudes are described by functions that are analytic away from the usual physical poles and cuts. The point is that analytic functions always attain their maximum at a boundary of their domain of definition.  In the context of scattering amplitudes, these boundaries are the cuts generated by multiparticle intermediate states. For \textit{physical kinematics} the amplitude along the cut is constrained by the conditions that probabilities add up to one -- i.e. by unitarity.  For this reason we focus on the two body scattering of the lightest particle in the theory since then all the usual cuts of the amplitude correspond to physical kinematics. In $1+1$ dimensions where unitarity can be directly applied at the level of the S-matrix (simply, $|S(s)|\le1$ for $s$ along the cuts) we are faced with a clean problem in the theory of complex functions of a single variable. As we have an analytic function on a domain with a boundary along which it is bounded, so we are able to constrain its values inside this region and in particular the various physical couplings which we define as residues of factorization poles. {Section \ref{sec2} contains a derivation of the two dimensional bound which is a significant refinement of that in \cite{paperII}.}

In this paper we move the focus to higher dimensions which contains a plethora of very interesting and difficult elements absent in the simpler $1+1$ dimensional case. An essential difference is that the most convenient way to formulate unitarity requires introducing {\it partial waves} and these are not bounded by unitarity along their entire boundary (only along the so-called ``right cut''). Therefore the simple complex analysis argument of $1+1$ dimensions cannot directly apply. Furthermore, the analyticity and crossing symmetry requirements involve the \textit{amplitudes} rather than the partial waves, which forces one to use both descriptions of the scattering event. Still, it is possible to overcome these technical obstacles. We shall introduce a kind of uniformization coordinates where the full space of physical kinematics is mapped to (a few) unit circles. This will allow us to Taylor expand the amplitudes in a convergent and manifestly crossing symmetric way in the full physical plane and then to numerically impose unitarity along the physical boundaries.

We start by revisiting the two dimensional results with this new approach in section~\ref{sec2}, setup the higher dimensional problem in section~\ref{sec3} and present and analyze the corresponding numerical results in section~\ref{sec4}. In section \ref{sec:bonus} we compare our numerical results with the completely orthogonal approach of \cite{paperI} which is based on QFT in AdS and in particular does not require any analyticity assumptions. We conclude in section~\ref{secConc}. A number of appendices are included to complement the main text presentation.

\section{Two Dimensions Redux and Unit Circles}  \la{sec2}

In this section we revisit the much simpler two dimensional problem. In two dimensions we can solve things analytically, and so it is a great training ground for developing intuition and testing any new numerical approaches. Nonetheless, for the braver readers eager to learn about the higher dimensional story, this section can be skipped without compromising the logic of the paper. 

Most of the mathematical analysis of \cite{paperII} boils down to minor variations of the following simple problem: 

\begin{itemize}
\item[Q:] Consider all real analytic functions $f(z)=\left[f(z^\star)\right]^\star$ with no singularities inside the unit disk apart from a simple pole at $z=0$ and which are bounded on the unit circle as $|f(e^{i \phi})|\le 1$. 
\footnote{In addition, $f(z)$ should not have an essential singularity at the boundary of the disk such that $|f(z)|$ diverges as we approach the boundary from any direction inside the disk.}
What is the maximum possible residue at $z=0$ and which function has that residue? 

\item[A:] The maximum residue is~$1$ and the corresponding function is~$f=1/z$. 
\end{itemize}

\noindent Indeed~$g(z)=f(z)/(1/z)$ has no singularities inside the disk and obeys $|g(z)|\le 1$ at the boundary of the unit disk. By the so-called maximum modulus principle, it satisfies $|g(z)| \le 1$ \textit{everywhere} inside the disk. Its value at the origin -- which is nothing but the residue of~$f$ -- is therefore at most $1$. This maximum value is attained when $g$ is constant everywhere, that is when $g(z)=1$ corresponding to~$f(z)=1/z$. 
\begin{figure}[t]
\begin{center}
\includegraphics[scale=1.3]{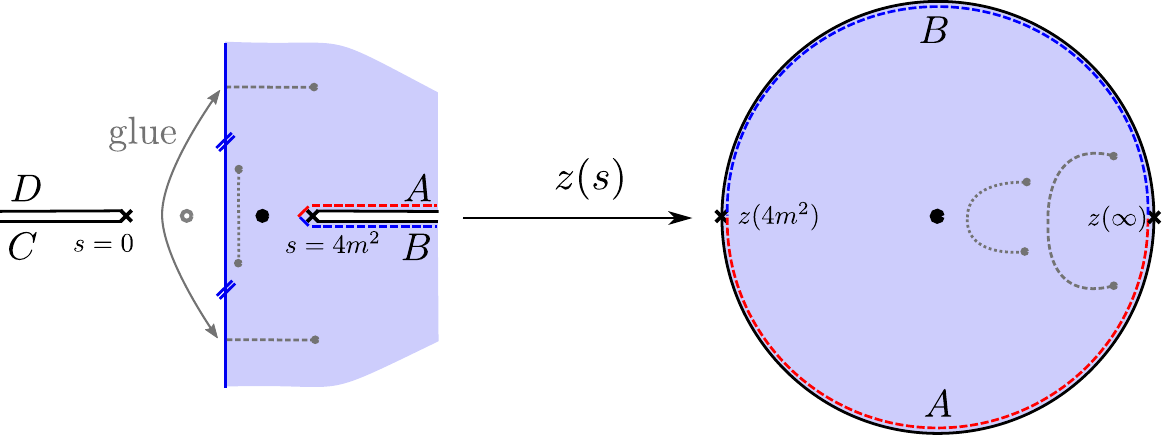} 
\end{center}
\caption{Mapping from the cut $s$-plane to the unit disk given in equation \eqref{cut_plane_to_disk}.  The mapping associates the points  $z(2+i y)=z(2-i y)$ and maps the half plane $\text{Re}(s) > 2$ to the full unit disk.  The grey, dashed curves on the left map to those on the right and are included to help the reader visualize the mapping.
}\label{cut_plane_to_disk_fig}
\end{figure}

To see how this simple problem relates to the analysis in \cite{paperII,Creutz} consider the $2\to2$ S-matrix~$S(s)$ for scattering of identical neutral particles of mass $m$ considered in \cite{paperII}. Assume also that there is a single bound-state showing up in this S-matrix element and for simplicity assume its mass $m_{b}>\sqrt{2}m$. Because of crossing symmetry $S(s)=S(4m^2-s)$ and we can focus on the region $\mbox{Re}(s)>2m^2$ without any loss of generality. In this half plane we have a threshold cut starting at $s=4m^2$, the bound-state pole at $s=m_b^2$ and no other singularities. Consider then the change of variable 
\beq  \label{cut_plane_to_disk}
z = \frac{\sqrt{s(4m^2-s)}-m_b\sqrt{4m^2-m_b^2}}{\sqrt{s(4m^2-s)}+m_b\sqrt{4m^2-m_b^2}} \qquad \text{or} \qquad \Big(\frac{s-2m^2}{2m^2}\Big)^{\!\!2} \!= 1+\frac{m_b^2}{m^2}{  \Big(\frac{m_b^2}{4m^2}-1 \Big)} \Big(\frac{z+1}{z-1}\Big)^2
\eeq
which maps this half plane into the unit disk, the bound-state pole into the origin of that disk and finally the threshold cut -- where unitarity is to be imposed -- to the boundary of the disk, see figure \ref{cut_plane_to_disk_fig}. In terms of $z$ the S-matrix is therefore exactly constrained by the conditions of the previous point; it has a pole at $z=0$ and obeys $|S(z)|\le 1$ at the boundary of the disk.\footnote{Note that this condition also holds on the lower half of the disk due to real analyticity.} Its maximum residue -- which is where we measure the (square of the) coupling to the bound-state -- is therefore $1$ and the corresponding optimal S-matrix is therefore $S(z)=1/z$. 

To recover the results of \cite{paperII} -- see e.g. formula (36) therein -- we simply need to take into account the Jacobian to go from $z$ to $s$, the simple kinematical multiplicative factors relating the S-matrix and the T-matrix and a factor of $m^4$ to render the coupling dimensionless. All other results of \cite{paperII} for more complicated bound-state spectra can be treated through simple generalizations of this simple example!\footnote{Strictly speaking the map to the unit circle is not even needed here. It suffices to assume there is no essential singularity at infinity so that the unitarity cut is the boundary of the region where $S(s)$ takes values. Then $S(s)/z(s)^{-1}$ is free of singularities in the physical region and obeys $|S(s)|\le 1$ on the cuts which are the boundaries of this region. Hence it can at most be one inside by the maximum modulus principle and the bound on the residue of $S$ follows. This is the argument in \cite{Creutz}. We still found the unit circle discussion to be useful as a warm-up to the higher dimensional case.} 

Although redundant at this point, it is instructive for what will come next in higher dimensions to set up this exactly solvable problem numerically. We define a function $S(z)$ in the unit circle as a pole plus a convergent Taylor expansion which we truncate at some large power $z^M$. Then we simply maximize the residue with the constraint that in a tightly spaced grid of~$K$ points on the unit circle  unitarity is satisfied. In \texttt{Mathematica}, the simple code below does the job: \\ \\
\noindent
\verb"M=20; K=50;" \\
\verb"S[z_] = residue/z + Sum[c[n] z^n, {n, 0, M}];" \\
\verb"variables = {residue}~Join~Table[c[n], {n, 0, M}];"\\
\verb"constraints = Table[S[Exp[I x]] S[Exp[-I x]] <= 1, {x, 0, \[Pi], \[Pi]/K}];" \\
\verb"FindMaximum[{residue, constraints}, variables]"  \\

\noindent
This nicely yields $residue\simeq 1$ and $c_n \simeq 0$ with great numerical accuracy which can be always improved. The reader is encouraged to copy/paste this and try by him/herself. It should take about $2$ or $3$ seconds to run.

\begin{figure}[t]
\begin{center}
\includegraphics[scale=1.2]{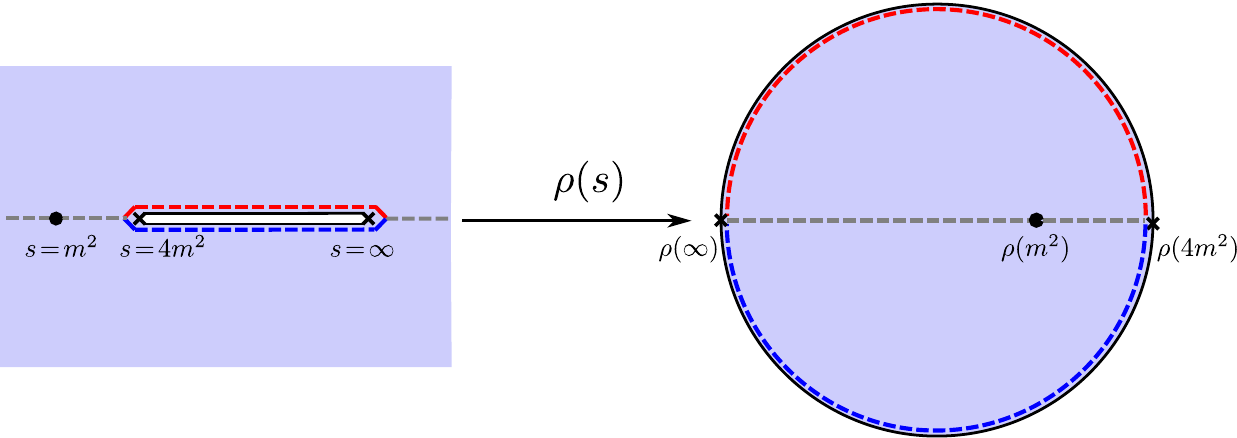} 
\end{center}
\caption{Mapping from the cut $s$-plane to the unit disk given in equation \eqref{rhoVar}.}\label{rhoMap}
\end{figure}

As a last warm-up it is very useful to solve this very same problem in a third way since this last approach is the closest to what we will do in higher dimensions. In this last approach to the problem we start by thinking of the S-matrix as being a function of both $s$ and $t$ as if they were independent variables; they are not since $s+t+u=4m^2$ and $u=0$ in two dimensions.\footnote{More precisely, either $u=0$ or $t=0$ corresponding to backward and forward scattering.} Then $S(s,t)$ is a function with a cut for $s>4m^{2}$, another cut for $t>4m^{2}$ as well as poles for single-particle processes in the $s$- and $t$- channels. Next we use a very convenient change of variable which maps the full complex plane with those cuts removed into the unit disk. This is the map 
\beq
s\mapsto \rho_s=
\frac{\sqrt{4m^{2}-s_0}-\sqrt{4m^{2}-s}}{\sqrt{4m^{2}-s_0}+\sqrt{4m^{2}-s}}\,,\qquad
\;\;\;\;\; s=\frac{s_0(1-\rho_s)^2+16m^{2}\rho_s}{(1+\rho_s)^{2}} \la{rhoVar} \,.
\eeq
where $s_0<4m^2$ is a free parameter that we can choose according to convenience. In the present case, it is convenient to choose $s_0=2m^2$ so that $\rho_s=0$ corresponds to the crossing symmetric point $s=t=2m^2$.
A similar map is also very useful in conformal bootstrap studies \cite{Hogervorst:2013sma}. It is illustrated in figure \ref{rhoMap}. The top of the cut maps to the upper boundary of the unit disk and the bottom of the cut maps to the lower boundary of the disk.  
The interval $\[0,4m^{2}\]$ maps to the interval $\rho \in \[2\sqrt{2}-3,1\]$ so this is where we find the poles associated to stable particles.

Apart from the poles corresponding to single particle exchanges, $S(\rho_s,\rho_t)$ is analytic for both $\rho_s$ and $\rho_t$ inside the unit disk and thus we can write
\beq
S(s,t) = -\frac{\hat g^2}{s-m_b^2}-\frac{\hat g^2}{t-m_b^2}+ \sum_{a,b=0}^{\infty} c_{ab}\, \rho_s^{a}\rho_t^{b} \la{repRho}
\eeq
Crossing symmetry is guaranteed provided the coefficients of the convergent Taylor expansion are symmetric,~$c_{ab}=c_{ba}$. Since we are going to evaluate the S-matrix on the constraint surface $s+t=4m^2$ we can simplify this ansatz further. In terms of $\rho_s$ and $\rho_t$ this constraint yields
\beq
\rho_s^2 \rho_t+\rho_t^2 \rho_s+4 \rho_s \rho_t 
+\rho_s + \rho_t= 0 \la{constRho}
\eeq
This means the representation (\ref{repRho}) has a big redundancy. 
We can always add to it polynomials in the left hand side of the constraint (\ref{constRho}). To remove this ambiguity, we can set to zero many of constants $c_{ab}$ (in appendix \ref{ap:rhoconstraints} we explained in detail which $c_{ab}$ can be set to zero).

\begin{figure}[t]
\begin{center}
\includegraphics[scale=0.8]{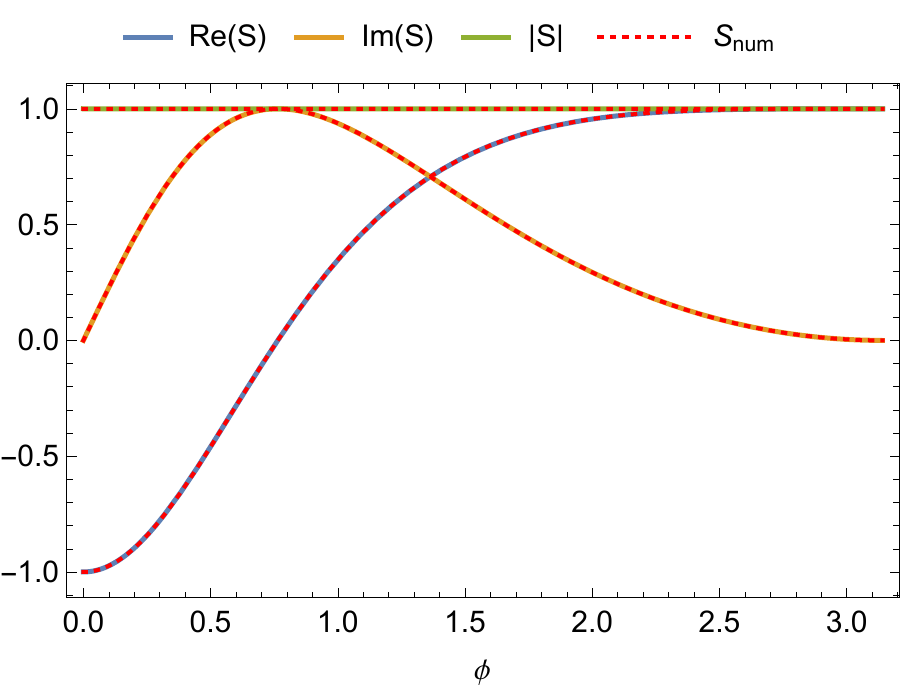} 
\end{center}
\caption{Comparison of the exact optimal S-matrix (given by $1/z(s)$ with $z$ given by \eqref{cut_plane_to_disk}) to numerical results using the ansatz \eqref{repRho} with the $a,b$ series truncated at maximum degree $N=5$ and $m_{BS}^{2}=3$.  We plot the physical region $\rho=e^{i \phi}$ with $\phi \in\[0,\pi\)$. The numerical results (red dashing) are indistinguishable from the exact results.}\label{S_num_comparison}
\end{figure}

Numerically, we set a cut-off in the sum (\ref{repRho}) and impose unitarity for $s>4$  which corresponds to the upper half circle where $\rho_s=e^{i\phi}$ with $\phi\in\[0,\pi\]$.  We evaluate $|S(s,t)|^2$ in a uniform grid in the $\phi$ interval which gives a set of quadratic constraint equations on the $c_{ab}$ and the residues of the poles.  We optimize $\hat g^{2}$ in the usual way using \texttt{FindMaximum} for example. The outcome of this third approach is in perfect agreement with our previous analytical and numerical results as illustrated in figure \ref{S_num_comparison}.  

To summarize: In two dimensions we can find the optimal S-matrix with largest possible residue analytically. 
\footnote{\label{essentialsingularitycounterexamplefootnote}
Notice that if we allow essential singularities at $s=\infty$ then there is no upper bound on $\hat{g}^2$. To see that consider the ansatz
\beq
S(s,t)=-\frac{\hat{g}^2}{s-m_b^2} \exp{\left[-\left(\frac{s-m_b^2}{\delta m^2} \right)^{2n} \right]} +(s\leftrightarrow t)\,,
\eeq
where $\delta m^2 = \frac{1}{2} \min (4m^2-m_b^2,m_b^2)$. For any value of $\hat{g}^2$, we can find a (large) positive integer $n$ such that this ansatz satisfies the unitarity constraint $|S(s,4m^2-s)| \le 1$ for $s>4m^2$. 
We thank Etienne Granet for raising this point.
We exclude such essential singularities at $s=\infty$ because they are incompatible with causality (see for instance appendix D of \cite{Camanho:2014apa}).
}
We do so by dividing the S-matrix by a clever guess and using the maximum modulus principle to show that this ratio should be one.  We recovered the same analytic results numerically in two ways. In the first one we start from a parametrization of the kinematics where we can Taylor expand the S-matrix and then truncate that expansion to obtain a finite algebraic problem which we can put on a computer. The second numerical approach is a small variation where we think of the S-matrix as a function of $s$ and $t$ as if they were independent and then consider a double Taylor expansion in each of them. 

What we implicitly used in the last method can be called an analytic \textit{extension} -- note that it is \textit{not} an analytic \textit{continuation} as we are increasing the number of variables and not just moving into the complex plane keeping the number of variables fixed. In this extension we promoted the S-matrix to a more general function of two variables which has no singularities in the cut $s$ and $t$ planes.\footnote{Of course we still have the poles associated with stable particles but these can be easily treated separately as in \eqref{repRho}. Here, we focus on the parametrization of the analytic part of the S-matrix.}
Equivalently, in terms of the $\rho$ variables, we assumed the existence of an extension into a function $S(\rho_s,\rho_t)$ which has no singularities in the polydisk $\{\rho_s,\rho_t \text{ such that } |\rho_s|\le 1 \text{ and } |\rho_t| \le 1\}$ while all we know a priori is that such a regular function exists only in the intersection of the polydisk with the constraint (\ref{constRho}). Why do we have the right to assume that such an extension exists at all? For instance, it could happen that such an extension would inevitably introduce new singularities in the full polydisk domain which would then invalidate the convergence of the double expansion (\ref{repRho}). Numerically, using this extension method we seem to find perfect agreement with the analytic results so somehow we should be safe. Indeed, the polydisk is a so-called Stein manifold\footnote{The unit disk is an open Riemann surface and those are Stein manifolds. Products of Stein manifolds are also Stein so the polydisk is also Stein.} and the constraint (\ref{constRho}) is an holomorphic embedding and as such defines a submanifold inside the polydisk which is also Stein. As discussed in greater detail below, there is a rather remarkable mathematical result which states that regular analytic extensions from Stein sub-manifolds inside Stein manifolds to the full Stein manifold do exist! The perfect numerical agreement is thus to be expected. 

Of course, in two dimensions this discussion is a clear use of excessive force. On the other hand, in higher dimensions we will also make use of such analytic extensions and there we will not have the luxury of the analytic results to cross-check our numerics. The theorem alluded to above generalizes to that case as well and is key in providing confidence for the higher dimensional numerics. 

There is also another more pedestrian explanation of why the double Taylor expansion numerics had to work which we present in appendix \ref{appendixA}; however, contrary to the discussion above, it makes use of particular features of the two dimensional problem and is not that useful as a warm up to the higher dimensional case. 

\section{Higher Dimensions}  \la{sec3}
We now move on to scattering amplitudes in $d+1$ spacetime dimensions. Consider again the elastic scattering process of two identical real scalar particles of mass $m$. In our conventions the S-matrix element is
\beq \label{eq:Smatrixel}
\< \bp_3, \bp_4 | S | \bp_1, \bp_2  \> = {\mathds 1} + i(2\pi)^{d+1} \delta^{(d+1)}(p_1 + p_2 - p_3 - p_4) M(s,t,u)
\eeq
with normalization such that 
\beq
{\mathds 1} = (2 \pi)^{2d} 4 E_{\bp_1} E_{\bp_2} \left( \delta^{(d)}(\bp_1 - \bp_3) \delta^{(d)}(\bp_2 - \bp_4)  + (3 \leftrightarrow 4)  \right) 
\eeq
where $E_{\bp}=\sqrt{m^2+\bp^2}$. The Mandelstam invariants are given by
\beq
s = (p_1 + p_2)^2 \qquad \qquad t = (p_1 - p_3)^2 \qquad \qquad u = (p_1 - p_4)^2
\eeq
which of course obey $s + t + u = 4 m^2$, and we henceforth work in units such that $m^2 = 1$. We often write $M(s,t) \equiv M(s,t,4 - s - t)$. In the channel under consideration $s$ is the squared center-of-mass energy and the scattering angle is given by
\beq
x = \cos(\theta) = 1 + \frac{2 t}{s - 4} = - 1 - \frac{2 u}{s - 4}
\eeq
Physical values of the Mandelstam invariants are therefore $4 \leq s$ and $4 - s \leq t \leq 0$. We can project onto channels with definite angular momentum by introducing the \emph{partial amplitudes}:
\beq
\label{eq:partialwaves}
S_\ell(s)= 1+ i \frac{(s-4)^\frac{d-2}{2}}{ \sqrt{s}}   \int\limits_{-1}^1\! dx\, (1-x^2)^{\frac{d-3}{2}}P^{(d)}_\ell(x) \left. M(s, t)\right|_{t\to \tfrac{1}{2}(s-4) (x-1)}
\eeq
where $P_\ell^{(d)}(x)$ is proportional\footnote{In general spacetime dimension, we have
$$P_\ell^{(d)}(x)=
\frac{l! \,\Gamma (\frac{d-2}{2} )}{4 (4\pi)^{\frac{d}{2}} \Gamma(d+l-2)} C^{(d-2)/2}_\ell(x)\,.
$$}
 to the Gegenbauer polynomials. In our conventions,
\beq
P^{(3)}_\ell(x) =\frac{1}{32\pi}   P_\ell(x)\,, \qquad \qquad  P^{(2)}_\ell(x) = \frac{1}{8\pi}  
\cos(\ell \theta)\ ,
\eeq
with $P_\ell(x)$ the usual Legendre polynomials, normalized such that $P_\ell(1) = 1$. We note that $S_\ell(s) = 1$ for odd $\ell$ because Bose symmetry implies invariance under the reflection $\theta \to \pi - \theta$.

Although the S-matrix element \eqref{eq:Smatrixel} has all kind of distributional properties, the amplitude $M(s,t,u)$ is a regular function (see e.g. \cite[section 4.3]{Weinberg:1995mt}). We will assume that $M(s,t,u)$ obeys three further constraints:
\begin{itemize}
  \item {\bf Crossing Symmetry}: $M(s,t,u)$ is completely symmetric in its arguments. The symmetry $u \leftrightarrow t$ follows from the aforementioned Bose symmetry, but the other generator of the crossing symmetry group can only be found from a more sophisticated analysis and requires the LSZ prescription.
  \item {\bf Analyticity}: $M(s,t,4 - s -t)$ is analytic for arbitrary complex $s$ and $t$, except for potential bound-state poles at $s = m_b^2$ with $0 < m_b^2 < 4$, a cut along the real axis starting at $s = 4$, and the images of these singularities under the crossing symmetry transformations. It further obeys the usual reality condition $M(s^*, t^* 4 - s^* - t^*) = M^*(s,t,4-s-t)$. We note that the analyticity assumption is actually rather optimistic, since this `maximal' analyticity has not been proven from axiomatic field theory.\footnote{%
  Certain analyticity properties are known to be valid very generally, derived either to all orders in perturbation theory or from axiomatic field theory; the latter case sometimes requires the Wightman axioms and other times merely requires the validity of the LSZ prescription and causality. Typically one can prove two-variable analyticity for all $s$ (modulo the known poles and cuts) but only for some finite range of values of $t$ or of $x$ which in particular includes the physical values. A standard result is that the proven analyticity is sufficient to analytically continue the amplitude from the $s$-channel to the $t$ or $u$ channels, establishing crossing symmetry \cite{Bros:1965kbd}. We refer to \cite{Martin:944356,Bjorken:1965zz} and references therein for more extensive discussions.
  }
  On the other hand some a posteriori justification is provided by the remarkable agreement between some of our results and those obtained without maximal analyticity in the older literature. We therefore believe that this assumption is sufficiently mild to generate physically meaningful results. We offer some further comments on this point in section \ref{sec:bonus} and the conclusions section below.
  \item {\bf Unitarity}: From $S^\dagger S = 1$ we find that the unitarity constraint for elastic scattering takes the form
    \beq\label{eq:unitarity}
  \left|S_\ell(s) \right| \leq  1 
  \eeq
  for all $s \geq 4$ and $\ell \in \{0,2,4,\ldots\}$. Generically no other channels are available for a finite window of values of $s$, starting at 4 and ending at a higher threshold (like $s = 9$ for three-particle scattering). In such a window the above inequality should in fact be saturated. In this work we will  not impose such saturation, but our numerics in principle allows for it.
\end{itemize}
The aim of the S-matrix bootstrap program (as we envisage it) is to use these general conditions to obtain concrete constraints on the behavior of the function $M(s,t,u)$ or the partial amplitudes $S_\ell(s)$ at interesting points. Many results from the previous century can be found in the textbook \cite{martinbook} and the reviews \cite{tourdefrance,Martin:944356}.

The recent works \cite{Caron-Huot:2016icg,Sever:2017ylk}  pursue a bootstrap analysis of scattering amplitudes of weakly interacting higher spin theories, where the amplitudes are meromorphic functions of the Mandelstam invariants. Analytically, they beautifully explore the large $s$ and $t$ regime of weakly interacting higher spin scattering amplitudes and observe remarkable universality there. In contrast, our analysis is fully non-perturbative and the only poles of the scattering amplitudes are associated with stable particles (below the 2-particle continuum). Nevertheless it would be very interesting to investigate the same large $s$ and $t$ regime within our numerical approach.

\subsection{Ansatz}
In this subsection we explore the consequences of our analyticity assumption in some detail. As a toy model we can start with a single-variable function $f(z)$ which is analytic in a simple domain $D \subset \mathbb C$. If we define $\rho : D \to \Delta$ as a biholomorphic map between $D$ and the unit disk $\Delta = \{ \rho \in \mathbb C \,: \,| \rho | < 1\}$, then any such $f(z)$ has a Taylor series expansion of the form
\beq
f(z) = \sum_{n = 0}^\infty c_n \rho(z)^n
\eeq
which converges as long as $|\rho(z)| < 1$. Our multi-variable problem is unfortunately not so easy, since for $M(s,t)$ the moving cuts imply that the domain of analyticity in one variable, say $s$, depends on the other variable $t$. We will remedy this as follows. First we relax the constraint $s + t + u = 4$ and consider three-variable functions $M(s,t,u)$. Then we transform the variables $(s,t,u) \to (\rho_s,\rho_t,\rho_u)$ using the map \eqref{rhoVar} which is, with $m^2 = 1$,
\beq\label{s_to_rho}
s\mapsto \rho_s=
\frac{\sqrt{4-s_0}-\sqrt{4-s}}{\sqrt{4-s_0}+\sqrt{4-s}}\,,\qquad
\;\;\;\;\; s=\frac{s_0(1-\rho_s)^2+16\rho_s}{(1+\rho_s)^{2}}\,.
\eeq
In this case, it is convenient to choose $s_0=\frac{4}{3}$ so that $\rho_s=\rho_t=\rho_u=0$ corresponds to the crossing symmetric point $s=t=u=\frac{4}{3}$.
Now, since the transformation $\rho_s$ maps the $s$-plane minus the right cut starting at $s = 4$ to the unit disk, we see that in the $\rho$ variables all the cuts lie \textit{outside} the polydisk $\Delta^3$ defined by $|\rho_s| < 1$, $|\rho_t| < 1$ and $|\rho_u| < 1$. The only remaining singularities are then the poles and it is natural to write
\beq \label{eq:Asum}
M(s,t,u)  = - \frac{g^2}{s-m_b^2}-\frac{g^2}{t-m_b^2}-\frac{g^2}{u-m_b^2} + \sum_{a,b,c=0} \a_{abc}\, \rho_s^a \rho_t^b \rho_u^c 
\eeq
where the triple $\rho$ series converges inside $\Delta^3$, and for definiteness we have put in the poles for a single scalar bound state of mass $m_b$. The demands of crossing symmetry are implemented by demanding that the coefficients $\a_{abc}$ are totally symmetric in their indices. When restricted to the surface defined by $s + t + u = 4$ the ansatz \eqref{eq:Asum} obeys the analyticity and crossing symmetry constraints. It is perhaps more surprising that the converse is also true: any function obeying the analyticity constraints on the surface $s + t + u = 4$ can be extended to a function on $\Delta^3$, analytic modulo the poles, and therefore can be written in the form \eqref{eq:Asum}. This follows from a mathematical theorem known as Cartan's theorem B, which is a statement about the vanishing of higher cohomologies of coherent analytic sheaves on Stein manifolds (see e.g. \cite{forstnerivc2011stein}) -- in the case at hand this implies that there is no obstruction to an extension away from the surface $s + t + u = 4$.\footnote{In contrast to the Mandelstam representation, notice that our ansatz \eqref{eq:Asum} `solves' the constraints of analyticity and crossing symmetry without demanding specific asymptotic behavior for large values of the Mandelstam invariants. We offer more comments on the relation between our ansatz and the Mandelstam representation in appendix \ref{Man}.}

The triple $\rho$ expansion in equation \eqref{eq:Asum} is the starting point for our numerical work. Our approach is to restrict the expansion to a finite sum by imposing
\beq
a + b + c \leq N_{\max}
\eeq
and then further restricting to the constraint surface $s + t + u = 4$ which is given by a polynomial equation
\beq
\rho_s^2 \rho_t^2 \rho_u 
+\rho_s^2 \rho_u^2 \rho_t +\rho_t^2 \rho_u^2 \rho_s + \text{(lower degree terms)} = 0
\eeq
and which in practice allows us to eliminate many terms 
in  \eqref{eq:Asum} (in appendix \ref{ap:rhoconstraints} we explain in detail which terms can be set to zero).
The remaining freedom in our ansatz then consists of the finitely many remaining $\alpha_{abc}$ together with the bound state parameters; since this is a finite-dimensional space we can use a computer to numerically explore the space of scattering amplitudes. Of course we want to keep $N_{\max}$ as large as possible. As we will see, in fortunate cases the numerical results stabilize already for feasible values of $N_{\max}$, while in other cases we can extrapolate.\footnote{As discussed further in appendix \ref{subapp:largeenergy}, the unitarity constraints imply that the large energy behavior is somewhat restricted if we keep $N_{\max}$ finite, but we do not expect this to affect the physics in our results.}

It will be the job of the computer to impose the unitarity constraints, which are quadratic constraints in the parameters $g^2$ and $\alpha_{abc}$. Rather than checking the infinity of constraints for all $s$ and $\ell$, we impose a cutoff and check that unitarity constraints are obeyed only for $\ell \leq \ell_{\max}$ and along a grid of values for $s$. Experimentally we observe that our results remain meaningful if $\ell_{\max}$ is not much smaller than $N_{\max}$  
and if the grid is sufficiently refined. In appendix \ref{app:numerics} we discuss the dependence on these parameters in more detail, and outline the numerical implementation.

\section{Results} \la{sec4}

In this section we present our numerical results for several maximization problems using the S-matrix bootstrap method explained above.
For most of this section we restrict our attention to 3+1 dimensional QFTs, i.e. $d=3$ in our notation.
In the final subsection \ref{sec:bonus}, we consider $2+1$ dimensional QFTs.

\subsection{Cubic coupling}
\label{subsec:maxcoupling}
For our first result we consider a scattering amplitude with a single pole corresponding to the exchange of a scalar particle of mass $m_b$, exactly as in our ansatz \eqref{eq:Asum}, and maximize the value of the residue $g^2$ as a function of $m_b$.\footnote{For $m_b \neq m$ this in particular implies that there is by assumption no three-point coupling where all particles have mass $m$. This could be due to a symmetry but we do not have to commit to an underlying mechanism here.}

\begin{figure}
\begin{center}
\includegraphics[width=0.8 \linewidth]{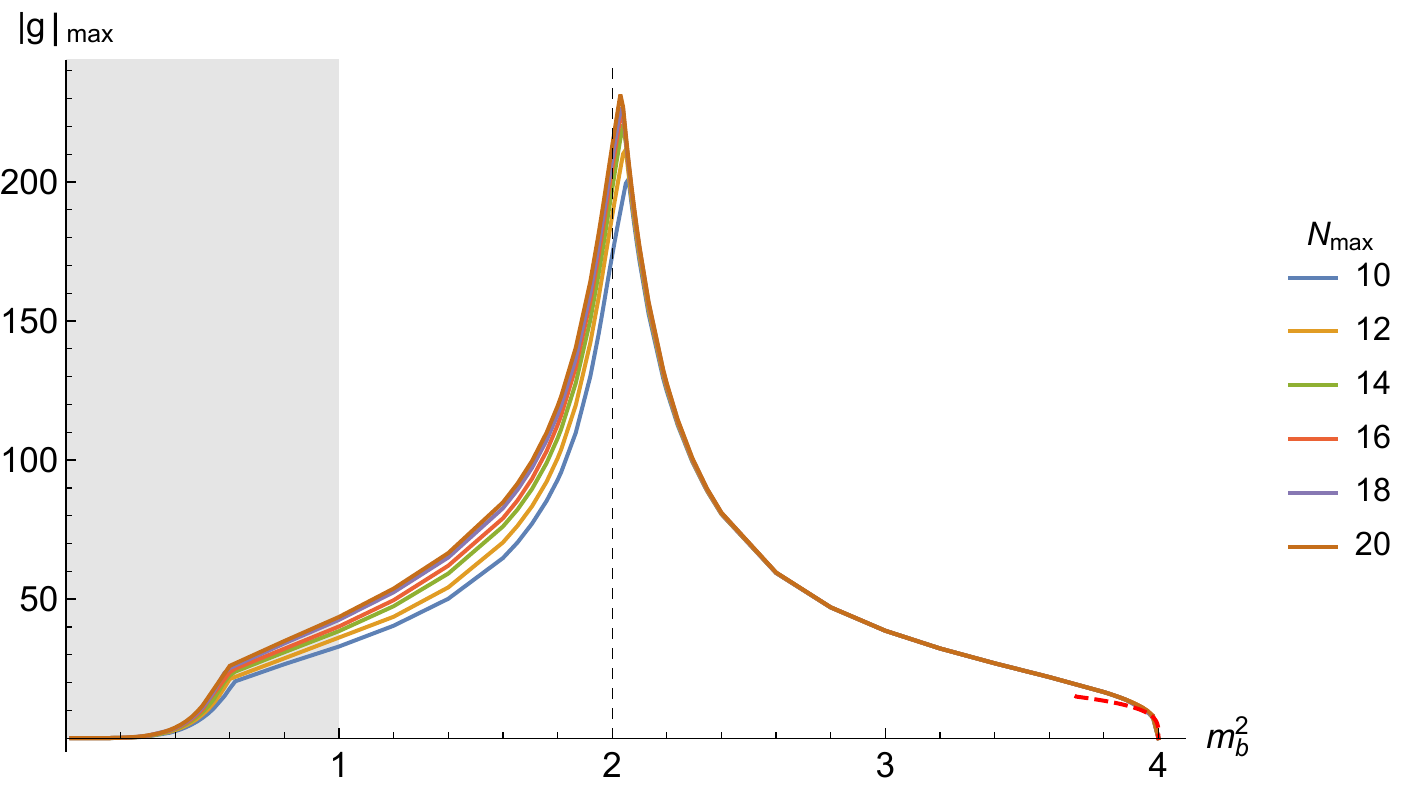}
\caption{\label{fig:upperbound_residue_4d}Largest possible value $|g|_{\max}$ as a function of $m_b^2$, using a triple rho expansion of the amplitude for the given values of $N_{\max}$ and after imposing the unitarity constraints for spins up to $\ell_{\max}=20$. As explained in the text, the shaded area is physically incompatible with our analyticity assumption. We added the analytic result of appendix \ref{NR} as the dashed line near $m_b^2 = 4$.}
\end{center}
\end{figure}

\begin{figure}
\begin{center}
\includegraphics[width=0.5 \linewidth]{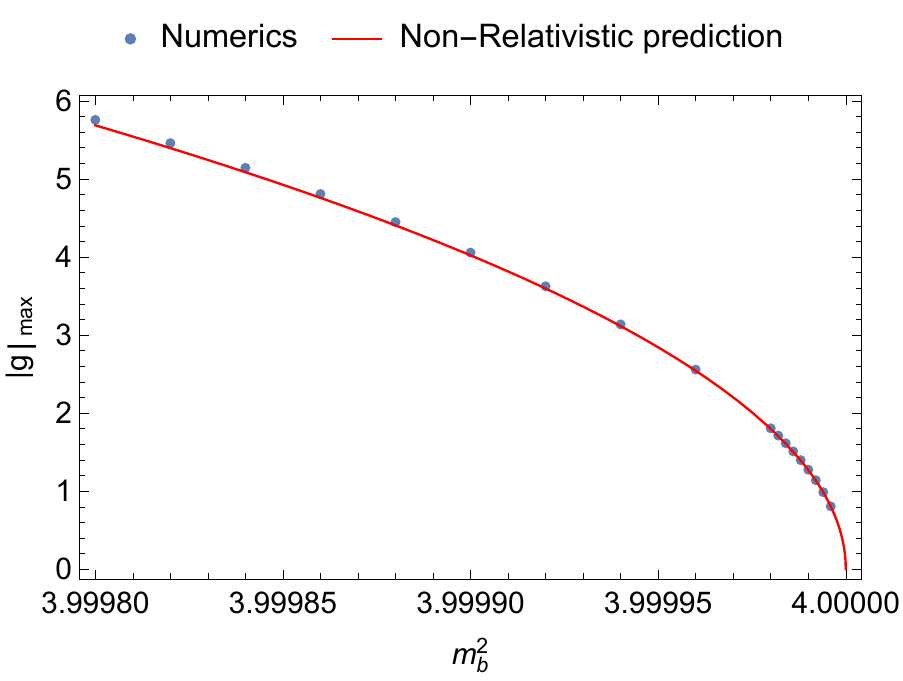}
\caption{\label{fig:non_rel_comparison} Comparison of numerics with the non-relativistic prediction $|g|_{\text{max}}\sim 256 \pi \sqrt{2 - m_{b}/m}$ derived in appendix \ref{NR}.  These numerics were performed with $s_{0}=m_{b}^{2}$ so that the bound state pole always maps to the centre of the $\rho$ disk.  This greatly expedites the convergence in this limit.  For example one can already see convergence with $N_{\text{max}}=2$ and $\ell_{\text{max}}=4$ which are the parameter values used for this plot.}
\end{center}
\end{figure}

In figure \ref{fig:upperbound_residue_4d} we plot the maximum absolute value of the coupling $|g|$ defined as the residue of the pole, with the different curves corresponding to different values of $N_{\max}$. We have obtained this plot by maximizing $|g|$ for a sequence of values of $m_b$ and the indicated curve is an interpolation through our data points. The plot is rather rich; we discuss its key features one by one.
\begin{itemize}
  \item Convergence with $N_{\max}$. For $m_b \gtrsim \sqrt{2}$ we see that $|g|_{\max}$ is nearly stationary as we vary $N_{\max}$, whereas for $m_b \lesssim \sqrt{2}$ we observe more significant improvements with $N_{\max}$. We have no explanation for this disparate behaviour (although we suspect it to be related to some subtler higher energy behaviour to which our ansatz is struggling to converge -- see also discussion section \ref{secConc} and appendix \ref{2Dexample}). Numerically we find that we can extrapolate to infinite $N_{\max}$ and appear to get a finite answer in either domain. We expect this value to correspond to an upper bound on $|g|$ for \emph{any} scattering amplitude that obeys the constraints of the previous section.\footnote{As for any of the results in this paper, it might very well be possible to derive even stronger bounds by including the constraints from other processes involving more particles.}
  \item Peak near $m_b \sim \sqrt{2}$. The clear peak is reminiscent of two-dimensional scattering amplitudes, where it was easily explained because in that case the $s$- and $u$-channel poles cancel precisely at $m_b = \sqrt{2}$ and the number $|g|$ becomes meaningless -- so no upper bound can be obtained.\footnote{In our ansatz \eqref{eq:Asum} this is easily observed by recalling that $t = 0$ in two dimensions, so also $u = 4 - s$. There is also only one partial wave with $\ell = 0$.} In greater than two dimensions the cross-channel poles are smeared into a cut by the projection onto the partial waves.  One can easily see from \eqref{eq:partialwaves} that this cut starts at $s=4-m_{b}^{2}$ thus we find in the partial amplitudes the $s$-channel pole starts to overlap with the $t$- and $u$-channel cut when $m_{b}^{2}\le 2$.  While there is a singularity at the branch point of this cut with the correct sign to ``screen'' the $s$-channel pole, this singularity is not strong enough to fully cancel the pole as in $1+1$ dimensions.  The singularity is a $\log(s-4+m_{b}^{2})$ in $3+1$ and $(s-4+m_{b}^{2})^{-1/2}$ in $2+1$ (see appendix \ref{analyticIntegrals} for the expicit expressions). We thus expect the peak in figure \ref{fig:upperbound_residue_4d} to remain finite as $N_{\max} \to \infty$. This is borne out by some crude extrapolations (not shown).
  \item Behavior near threshold, $m_b \sim 2$. As explained in appendix \ref{NR}, when $m_b - 2$ is parametrically small we can analytically constrain the behavior of $|g|_{\text{max}}$ as a function of $m_b$. This result is plotted in the figure as the dashed red line segment.  Figure \ref{fig:non_rel_comparison} shows a closer analysis of this limit. We see that it accurately traces our numerical results, with the agreement improving as $m_b$ approaches $2$.
  \item Behavior for $m_b < 1$. In this region the scattered particle is no longer the lightest particle in the theory and on physical grounds we expect 
  the two-particle cut in $A(s,t,u)$ to begin at $2 m_b$ rather than at $2 m$. For small enough $m_b$ this is corroborated by our numerics since $|g|_{\max} \sim 0$ so no pole can be present without modifying our ansatz. It would be interesting to understand in more detail the kink near $m_b \approx 0.5 $.
\end{itemize}
For $m_b = 1$ we can identify the pole with an exchange of the external particle. Reference \cite{muller} (see also \cite{martinbook}) discusses an analytic upper bound on $|g|$ for that case which in our conventions takes the value:
\beq
|g| \lesssim 16 \pi \sqrt{1.5 \cdot 10^6} \approx 61562.4
\eeq
which is far weaker than our current bounds.\footnote{In \cite{martinbook} the author conceded that ``{\it [it] is a large number, but of course [the] calculation was only carried through to show that there \emph{exists} an upper bound.}'' We are however not aware of any better previous bounds in the literature.}

\subsection{Quartic coupling}
\label{sec:quartic}
\begin{figure}
\begin{center}
\includegraphics[width=0.6 \linewidth]{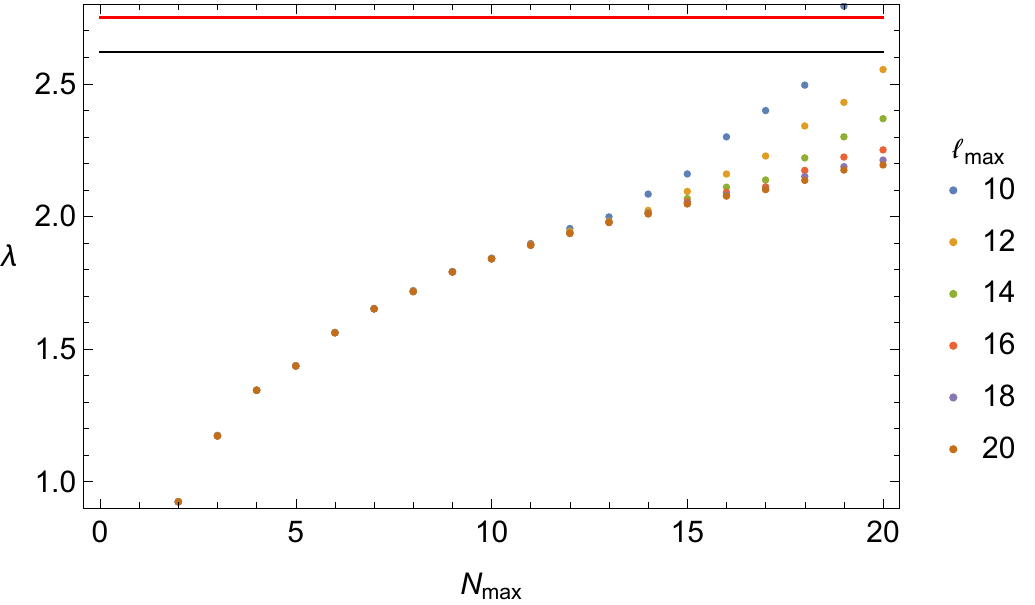}
\caption{\label{pion_upper_bound1}
A first attempt at obtaining a maximal value of the quartic coupling $\lambda\equiv \frac{1}{32\pi}M(\frac{4}{3},\frac{4}{3} ,\frac{4}{3} ) $, using the ansatz \eqref{eq:Asum} with $g=0$.
We impose the unitarity constraint \eqref{eq:unitarity} for all $\ell\le \ell_{\text{max}}$. Convergence requires larger $\ell_{\text{max}}$ for higher  values of $N_{\text{max}}$. With this ansatz, the maximal quartic coupling continues to increase significantly with $N_{\text{max}}$ even for $N_{\text{max}}=20$.  The black line indicates the value $2.262$ achieved in the solution of \cite{Auberson:1977ss}, while the red line indicates the rigorous upper bound $2.75$ of \cite{Lopez:1976zs}.  For large enough $\ell_{\text{max}}$ and $N_{\text{max}}$ our curves must eventually form a plateau between these two lines, however the convergence is so poor that this cannot be inferred from the plot.
}
\end{center}
\end{figure}  
Our second set of results concerns the scattering amplitudes $M(s,t,u)$ without any bound state poles, as for example would be the case in $\pi^0$ scattering. We will constrain the value of the amplitude at the symmetric but unphysical point $s = t = u = 4/3$ and therefore define:
\beq
\lambda\equiv \frac{1}{32\pi}M\left(\frac{4}{3},\frac{4}{3} ,\frac{4}{3} \right) 
\eeq
Historically $\lambda$ was taken to be a measure of the quartic pion interaction strength. In previous works \cite{Lopez:1976zs} it was constrained both from above and below, in our conventions:
\beq\label{eq:rigorous_bounds}
-8.2 \leq \lambda  \leq 2.75
\eeq
These constraints stem only from the use of axiomatially proven analyticity, crossing and unitarity. Another data point is provided by the explicit ``amplitudes'' constructed by Auberson and Mennessier, one with $\lambda = 2.62$ \cite{Auberson:1977ss} and one with $\lambda = -1.69$ \cite{Auberson:1979ye}, both of which obey analyticity, crossing and unitarity. This provides a lower bound for any upper bound and vice versa. It is particularly remarkable that there exists a fairly narrow interval $[2,62,2.75]$ in which the best upper bound must reside.

\begin{figure}
\begin{center}
\includegraphics[width=0.6 \linewidth]{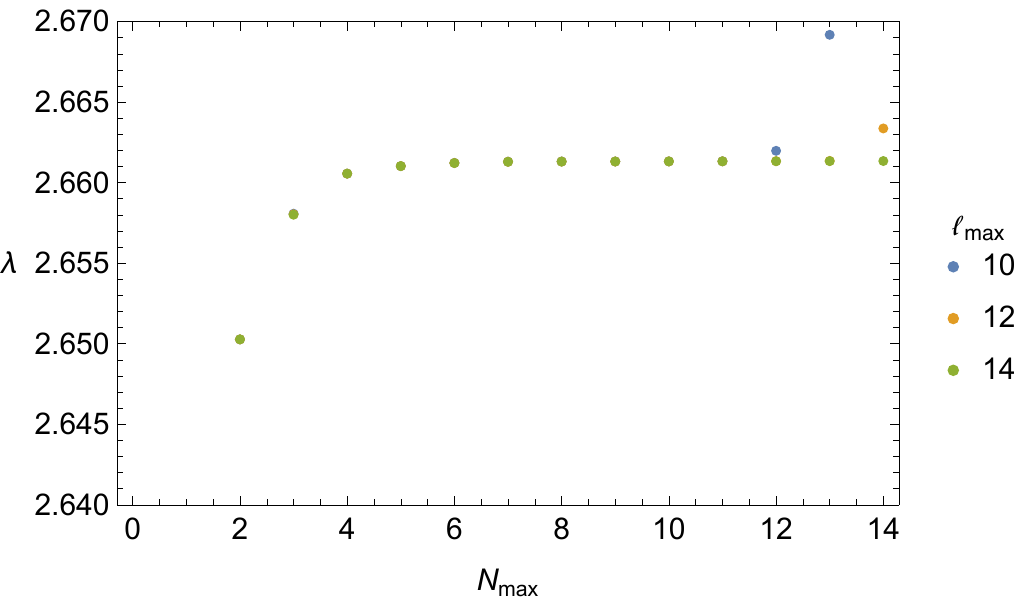}
\caption{ \label{pion_upper_bound}
Maximal value of the quartic coupling $\lambda\equiv \frac{1}{32\pi}M(\frac{4}{3},\frac{4}{3} ,\frac{4}{3} ) $, now using the ansatz \eqref{eq:Asum} with $g=0$, supplemented with the term \eqref{eq:theshSing}.
With this improved ansatz, the maximal quartic coupling effectively saturates for  $N_{\text{max}} \gtrsim 6$.  A few values of $\ell_{\text{max}}$ are shown to demonstrate that the value of the plateau is independent of this cutoff -- the data points for various $\ell_{\text{max}}$ are indistinguishable until around $N_{\text{max}}\gtrsim12$ where the plateau is lost for $\ell_{\text{max}}=10$ (this is just the usual loss of the plateau when $N_{\text{max}}$ becomes too large compared to $\ell_{\text{max}}$).}
\end{center}
\end{figure}  

Let us first discuss the case of the upper bound. Figure \ref{pion_upper_bound1} shows the largest possible value on $\lambda$ using the ansatz \eqref{eq:Asum} (with $g=0$).  One can see that the convergence with $N_{\text{max}}$ is quite slow which suggests the presence of a singularity near or on the boundary of the $\rho$ discs.  Indeed, as pointed out in \cite{Lopez:1976zs, Auberson:1977ss} the amplitude which achieves the upper bound naturally has a singularity of the form $(s-4)^{-1/2}$ corresponding to a bound state sitting  precisely at threshold.  Physically this is intuitive:  the positive sign of the amplitude corresponds to an {\it attractive} interaction.\footnote{For example in a non-relativistic approximation this would correspond to an attractive delta function potential \cite{Coleman:1976uz}.}  The situation in which the interaction is as attractive as possible without introducing new bound states occurs just at the point where a resonance is pulled all the way to the threshold.  Mathematically it is natural that to make the amplitude as big as possible at the symmetric point it should be made as big as possible at threshold. Figure \ref{pion_upper_bound} shows the bound on $\lambda$ with the threshold bound state included in the anstaz.  This amounts to adding 
\beq\label{eq:theshSing}
\alpha \(\frac{1}{\rho_{s}-1}+\frac{1}{\rho_{t}-1}+\frac{1}{\rho_{u}-1}\)
\eeq
to the ansatz \eqref{eq:Asum} where now $\alpha$ is another parameter to be varied. This singularity does not cause a violation of unitarity because it is canceled by the phase-space volume factor in \eqref{eq:unitarity}. More precisely,
we find that the $\ell=0$
partial amplitude
near threshold behaves like
\beq\label{eq:ThreshSingLB}
S_0(s) = 1 + \frac{1}{16\sqrt{6}\pi} \alpha + O(\sqrt{s-4})\,.
\eeq
and therefore
\beq
\label{eq:alphaunitarity}
-32 \sqrt{6} \pi \leq \alpha \leq 0\,.
\eeq
The unitarity constraints for the higher spin partial amplitudes do not lead to further restrictions on $\alpha$.

Once the threshold bound state \eqref{eq:theshSing} is included we find that convergence is now quite rapid as indicated by the plateau in figure \ref{pion_upper_bound} already seen at modest values of $\ell_{\text{max}}$ and $N_{\text{max}}$.  The height of the plateau is $2.6613...$ and since
\beq
2.62 < 2.6613... < 2.75.
\eeq
it falls beautifully below the rigorous bound of \cite{Lopez:1976zs} but above the solution constructed in \cite{Auberson:1977ss}. Given the flexibility of our anstaz we expect this value to represent the strictest possible bound that derives from unitarity, crossing and analyticity of a single amplitude.  

\begin{figure}[!ht]	
\begin{center}
\includegraphics[height=5cm]{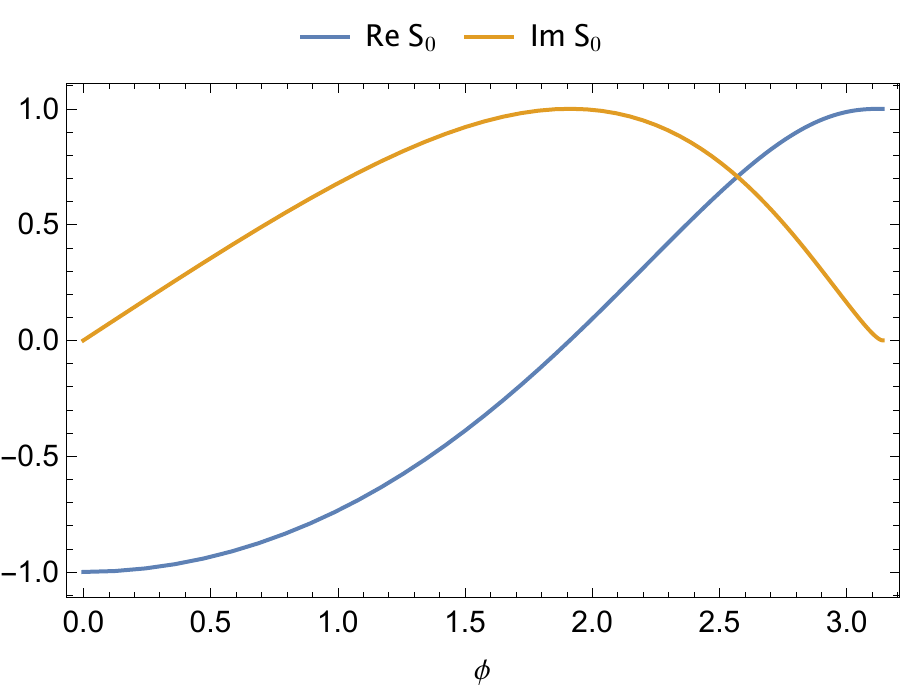}
\includegraphics[height=5cm]{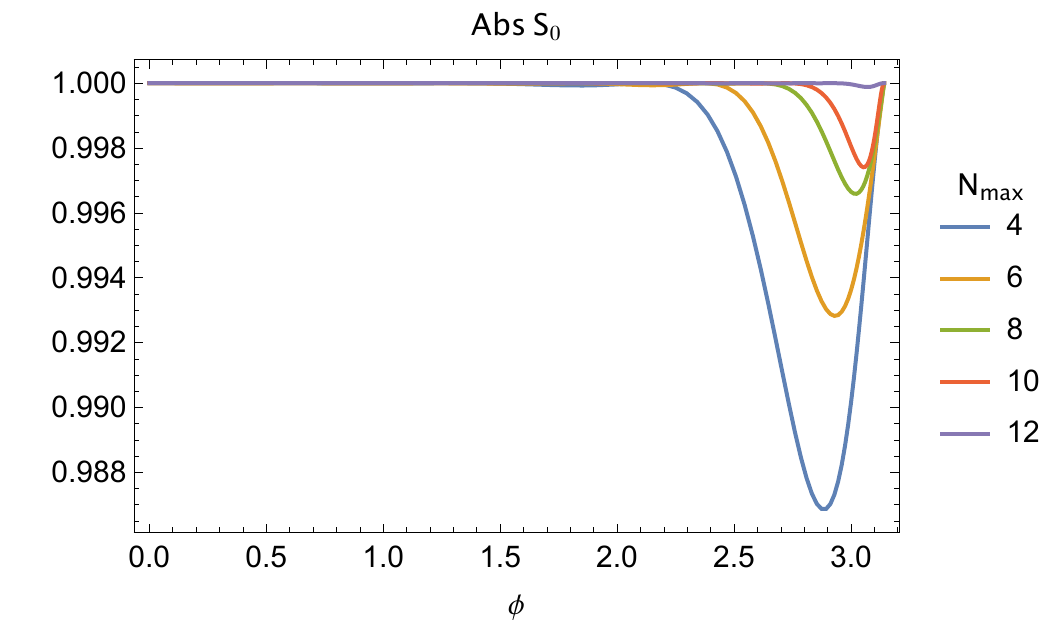}
\caption{\label{fig:no_particle_production} Real and imaginary parts of $S_{0}$ for $N_{\text{max}}=12$ and $\ell_{\text{max}}=20$ (left plot).  Absolute value of $S_{0}$ for $\ell_{\text{max}}=20$ and several values of $N_{\text{max}}$ (right plot).} 
\end{center}
\end{figure} 

An interesting feature of the optimal solution is what appears to be a tendency toward saturation of unitarity.  In right plot in figure \ref{fig:no_particle_production} one can see that $|S_{0}|$ increasingly saturates unitarity for increasing values of $N_{\text{max}}$.   
A related fact is that we observe numerically $\alpha = - 32 \sqrt{6}\pi$ to great accuracy indicating that unitarity is saturated at threshold.  Unitarity saturation is also observed in the higher partial waves.

\begin{figure}[!b]
\begin{center}
\includegraphics[width=0.6 \linewidth]{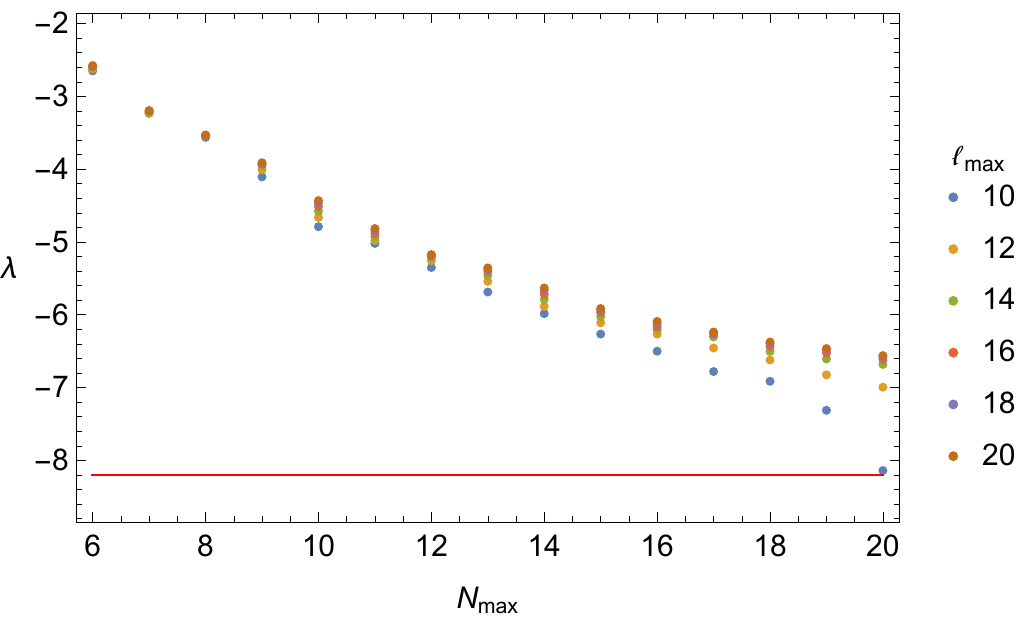}
\caption{\label{pi_lower_bound}
Minimal value of the quartic coupling $\lambda\equiv \frac{1}{32\pi}M(\frac{4}{3},\frac{4}{3} ,\frac{4}{3} ) $ achieved with the ansatz \eqref{eq:Asum} (with $g=0$).
With this ansatz, the minimal quartic coupling continues to decrease significantly with $N_{\text{max}}$ even for $N_{\text{max}}=20$. }
\end{center}
\end{figure}

Let us now consider the lower extremum for which our results are shown in figure \ref{pi_lower_bound}.  As in the previous case (without the threshold singularity) the convergence is quite slow in $N_{\text{max}}$.  Unfortunately the addition of a threshold bound-state of the form \eqref{eq:theshSing} cannot save us here, since we would need $\alpha > 0$ to lower the value of $\lambda$ but according to \eqref{eq:alphaunitarity} this is not allowed by unitarity of the spin 0 partial amplitude at threshold. Physically this makes sense -- if $\lambda < 0$ then this indicates a repulsive force which does not favour the creation of bound states nor moving resonances down to the threshold value. Unfortunately we were not able to identify the relevant singularity in this case and thus were not able to improve the slow convergence.

Notwithstanding these convergence issues, we did already significantly improve the lowest possible value of $-1.69$ that was explicitly constructed in \cite{Auberson:1979ye}. As the authors of that paper already noted, the discrepancy between their $-1.69$ and the lower bound $-8.2$ of \cite{Lopez:1976zs} means that either the lower bound is quite far from optimal, or that the behaviour of the amplitude which provides this bound is quite ``wild'' so as to not be contained within the space of functions they explored. Our results indicate that the latter scenario is the correct one since we do seem to be approaching a value in the ball park of the lower bound in \eqref{eq:rigorous_bounds}.
 
\subsection{Exploring scattering lengths}
Another set of observables that received interest in days long gone were the scattering lengths $a_\ell$. These are defined as the behavior of the partial waves when $s$ approaches its threshold value 4. We will restrict ourselves to four spacetime dimensions, i.e. $d = 3$, where it is typically defined as
\beq
a_\ell \colonequals \lim_{s \to 4} \frac{S_\ell(s) - 1}{i(s-4)^{\ell + 1/2}}\,.
\eeq
with the limit taken from above in order to make direct contact with experiment. The power of $s-4$ in the denominator arises as follows. One assumes that $\lim_{s \to 4} M(s,t)$ is finite for all $t$ in some neighborhood of zero. Analyticity in $t$ then allows one to write down a Taylor series expansion in $t$ whose radius of convergence remains strictly positive as $s \to 4$. Substituting $t = \frac{1}{2}(s-4)(x-1)$ and doing the $x$ integral in \eqref{eq:partialwaves} to project onto the partial waves of spin $\ell$ then gives a finite scattering length for all $\ell$ precisely with the given prefactor (recall that we are considering $d=3$). The factor of $i$ is included to make the scattering length real if $M(s,t)$ is real-analytic. In this section we will investigate constraints on the scattering length for amplitudes without bound state poles, so we will be using the ansatz \eqref{eq:Asum} without the pole terms.

Let us begin with the largest possible values of the scattering length. We first recall that, in ordinary quantum mechanics, scattering lengths are known to diverge when a resonance crosses the threshold value $s = 4$. In the $\rho$--variables in $d=3$ this can be seen by considering scattering amplitudes that locally take the form
\beq
- \mu \frac{P_\ell^{(3)}(x)}{\rho_s - 1 - \epsilon} + \ldots
\eeq
with the dots denoting subleading terms, which include permutations to make the amplitude crossing symmetric and other terms to make the amplitude unitary for $s$ away from 4. From unitarity near $s = 4$ we obtain the constraint
\beq
0 \leq \mu \leq \frac{2 \ell + 1}{(8 \pi)^2 \sqrt{4 - s_0}}
\eeq
where we recall that $s_0$ in our ansatz is equal to $4/3$ and we used that $\int_{-1}^1 dx\, P^{(3)}_\ell(x)^2 = [512 \pi^2 (2\ell + 1)]^{-1}$ in our conventions. The important observation here is that unitarity bounds $\mu$ independently of the value of $\epsilon$, whereas the contribution to the spin $\ell$ scattering length is given by
\beq
\frac{(16\pi)^2\mu}{(2\ell + 1)\epsilon}
\eeq
so by sending $\epsilon$ to zero from above we can get an infinitely large positive scattering length. Notice that $\epsilon < 0$ creates a pole on the physical sheet and this is disallowed by our ansatz.\footnote{In fact, for negative but small $\epsilon$ and $\ell = 0$ this amplitude reproduces precisely the extremal behavior for a bound state near threshold discussed in section \ref{subsec:maxcoupling} and in appendix \ref{NR}.}

\begin{figure}
\includegraphics[height=3.5cm]{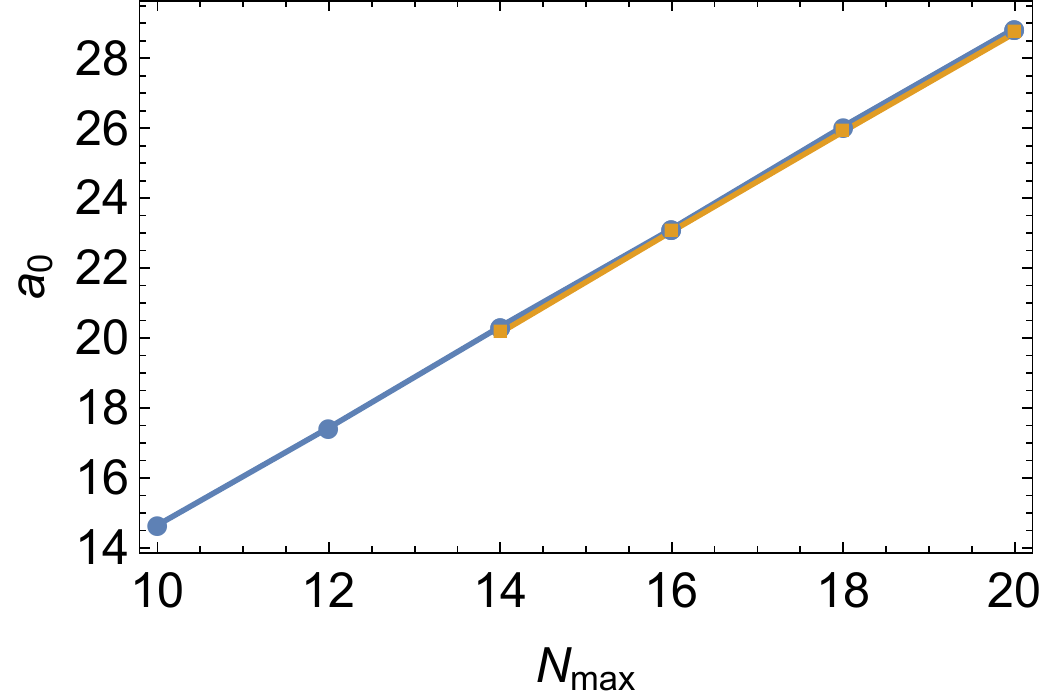}
\includegraphics[height=3.5cm]{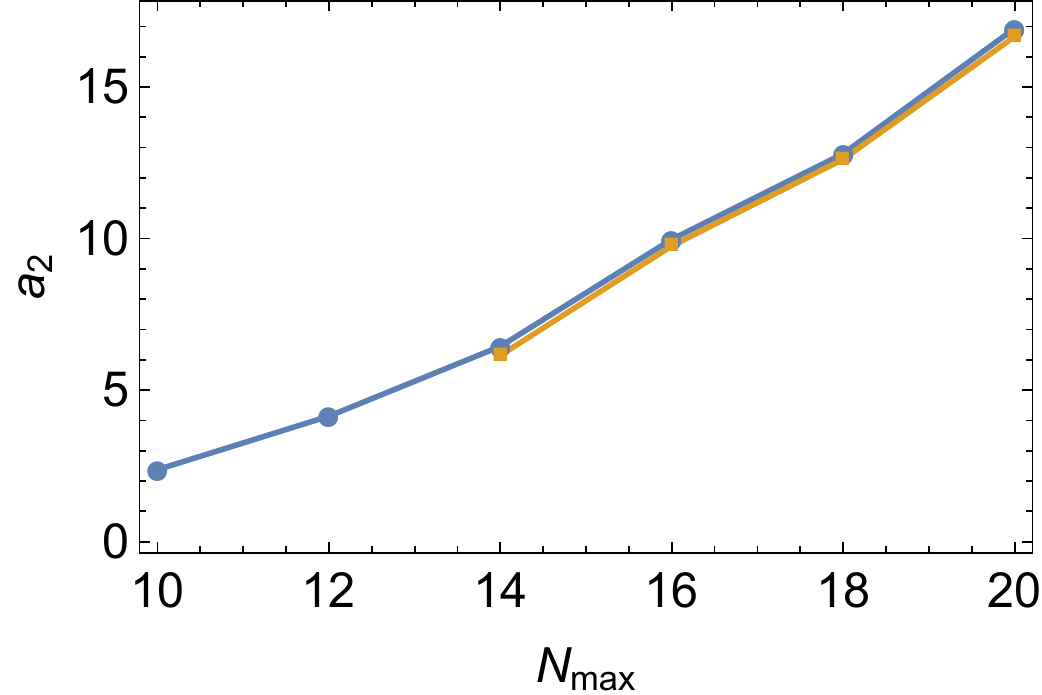}
\includegraphics[height=3.5cm]{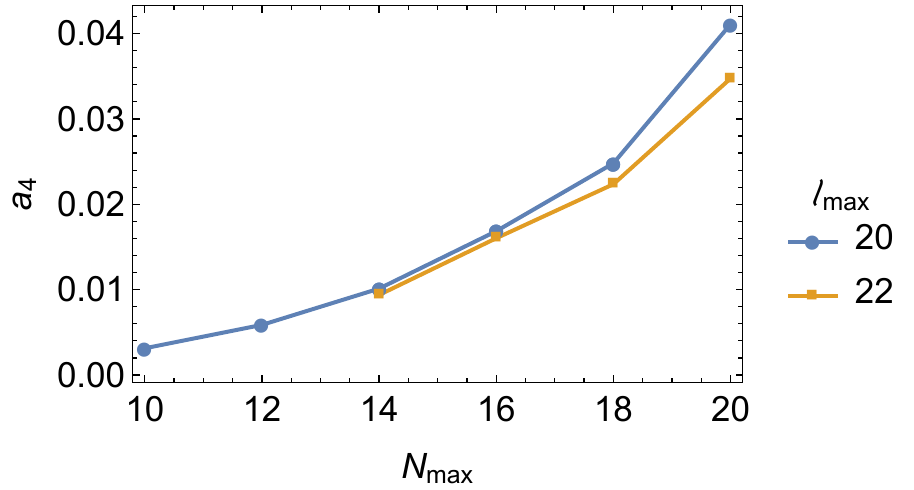}
\caption{\label{fig:largestscatteringlengths}Exploring large values of the scattering length. We plot the largest obtained spin 0, 2 and 4 scattering lengths as a function of the number of terms in our ansatz as parametrized by $N_{\max}$. For the larger values of $N_{\max}$ we include two values of $\ell_{\max}$. The results are in line with the expectation of unbounded growth as $N_{\max} \to \infty$.} 
\end{figure}

The unboundedness from above is borne out by our numerical results. In figure \ref{fig:largestscatteringlengths} we plot the largest possible values we can obtain for the spin 0, 2 and 4 scattering lengths with our usual ansatz \eqref{eq:Asum}, again with $g = 0$. We observe no convergence to a finite value as we increase $N_{\max}$.

\begin{figure}[h!]
\begin{center}
\includegraphics[width=0.6 \linewidth]{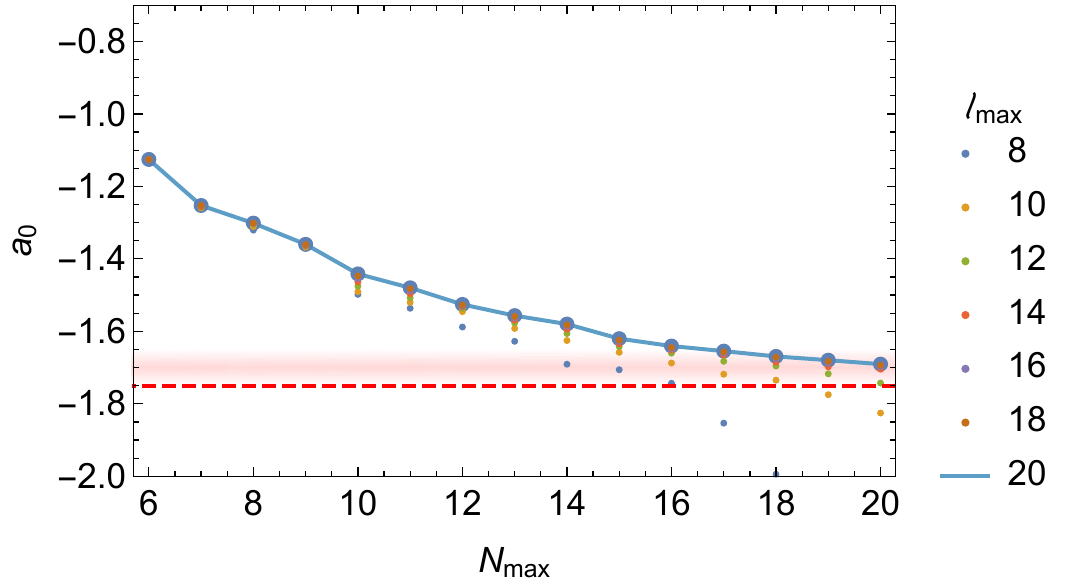}
\caption{\label{fig:smallestscatteringlength} Lowest possible value of the scattering length as a function of $N_{\max}$. Red dashed line: precise lower bound obtained in \cite{Lopez:1975ca}. Red shaded band: approximate lower bound obtained in \cite{Caprini:1980un}.}
\end{center}
\end{figure}

We can also consider the lowest possible values of the scattering lengths. For spin 0 the best known lower bound dates from 1980 and is given by \cite{Caprini:1980un}
\beq
a_0 \gtrsim -1.7\,,
\eeq
which slightly improves on a more precise bound obtained five years earlier in \cite{Lopez:1975ca}:
\beq
a_0 \geq -1.75\,.
\eeq
These result were the culmination of a series of works, starting with the observations in \cite{Lukaszuk:1967zz} which were followed by a series of intermediate improvements in e.g. \cite{martinbook,Bonnier:1968zz,Bonnier:1975xu,Caprini:1980un}.\footnote{Papers like \cite{Auberson:1979ye} contain a reference to an unpublished lower bound of -1.65 that had supposedly been obtained in 1978 by Caprini and Dita, the authors of \cite{Caprini:1980un}. It was confirmed to us by Irinel Caprini in personal communication that this value is incorrect.} Our numerical results are shown in figure \ref{fig:smallestscatteringlength} and are clearly converging in the neighborhood of the above lower bounds. This shows that the lower bound can more or less be saturated (with an amplitude that falls within our ansatz), which is actually a new result: the best known constructible value was -0.88 \cite{Auberson:1979ye}.

In fact, it may appear that we get dangerously close to the value $-1.7$ and that further increasing $N_{\max}$ may push us over the edge. However for this particular bound the convergence with $\ell_{\max}$ is quite slow and the value corresponding to infinite $\ell_{\max}$ may in fact increase a little bit. It would be interesting to perform a precision study with larger values of $\ell_{\max}$ and $N_{\max}$ and to simultaneously re-compute with higher precision the lower bound of $-1.7$ obtained in \cite{Caprini:1980un}. We leave this to the future.

For the higher spin scattering lengths one can use the Froissart-Gribov representation, see e.g. \cite{tourdefrance}, to arrive at the simple lower bound:
\beq
a_\ell \geq 0 \qquad \forall \ell \geq 2
\eeq
This is borne out by our numerics but we do not show the results since a plot consisting of nothing but zeroes is not very interesting.\footnote{We would like to remark that for sufficiently high $N_{\max}$ (say, 20) we need to impose unitarity for relatively large values of $\ell_{\max}$ (say, 24) before the lowest possible allowed value of $a_\ell$ gets pinned at zero.} 

\subsection{Bonus feature: three spacetime dimensions and QFT in AdS}
\label{sec:bonus}
In our previous work \cite{paperI} we outlined another method for constraining QFT data, based on putting a QFT in AdS. The main idea is to investigate the \emph{boundary correlation functions}, which behave exactly like CFT correlation functions (except there is no stress tensor) and are therefore amenable to an ordinary conformal bootstrap analysis. As we explained in \cite{paperI}, the translation between boundary and bulk quantities parallels the standard AdS/CFT dictionary, for example $m^2 R^2 = \Delta (\Delta - d)$, and furthermore we found precise formulae that dictate how the boundary correlation functions morph into flat-space scattering amplitudes upon sending the AdS curvature to zero. In \cite{paperI} we numerically tested these equations in 1+1 dimensions and found a quantitative match between the two approaches to the S-matrix bootstrap.

For this paper we set out to repeat this exercise for QFTs in 2+1 dimensions. We focused on the 2+1 dimensional version of the maximal possible coupling that we discussed in section \ref{subsec:maxcoupling}. This setup was called scenario I in \cite{paperI}. We discuss the salient points of the methodology before presenting the results.

\subsubsection{S-matrix bootstrap approach}
For the S-matrix bootstrap, the only difference in the implementation between the 3+1 dimensional analysis of section \ref{subsec:maxcoupling} and the present one is that we were no longer able to compute the partial amplitudes \eqref{eq:partialwaves} analytically.  The method explained in appendix \ref{analyticIntegrals} fails because the factor $(1-x^2)^{\frac{d-3}{2}}$ in \eqref{eq:partialwaves} introduces an additional square-root cut in $2+1$ dimensions ($d=2$ in the conventions of this paper).  Thus we are forced to evaluate the partial amplitudes by brute force use of Mathematica's \verb"NIntegrate".  Although slow, this approach is manageable with the use of multiple computing cores. This leads us to the:
\begin{itemize}
  \item[\color{red}\textbullet] {\color{red}First approach:} maximal three-point coupling $g^2$ for any flat-space QFT, obtained by assuming a flat-space scattering amplitude captured by our ansatz \eqref{eq:Asum} and obeying the unitarity condition \eqref{eq:unitarity}, as a function of $m_b/m$.
\end{itemize}

\subsubsection{QFT in AdS approach}
For the QFT in AdS approach we refer to \cite{paperI} for a detailed exposition of the method, except that presently we consider two-dimensional rather than one-dimensional conformal four-point functions. This implies that there is an extra cross ratio, since $z$ is no longer kinematically equal to $\bar z$, and conformal blocks are labelled by a pair $(\Delta,\ell)$ rather than just the scaling dimension $\Delta$. The combined effect of these modifications is simply that the numerics is computationally much more demanding.\footnote{The introduction of spin does lead to one new subtlety, namely the magnitude of the two-particle gap for spinning particles. If there is a single scalar particle corresponding to a boundary operator with dimension $\Delta$ then we chose to set the two-particle gap at $2\Delta + \ell$ as in free field theory. Notice that the flat-space limit merely dictates that the gap tends to $2 \Delta$ for very large $\Delta$, but there is freedom in choosing the subleading terms.}

Now, in \cite{paperI} we obtained a precise match in 1+1 dimensions by taking the raw numerical QFT in AdS results and performing a double extrapolation: first to ``infinite computational power'' and then to infinite $\Delta$ corresponding to the flat-space limit. For our 2+1 dimensional results we unfortunately run into trouble at the first step: our numerical results, obtained for $1 \leq \Delta \leq 20$ with functionals with up to 136 components, were not amenable to reliable extrapolations. We therefore chose to present the result \emph{directly for a QFT in AdS}. We chose $\Delta = 17$ as a representative value.\footnote{For $\Delta = 17$ we find that $m^2 R^2 = \Delta (\Delta - 2) = 255$ so the reduced compton wavelength of the particle is about 16 times the AdS radius of curvature in our setup - in this sense space is already quite flat.} Altogether this gives the:
\begin{itemize}
  \item[\color{mathematicablue}\textbullet] {\color{mathematicablue}Second approach:} maximal three-point (bulk) coupling $g^2$ for a QFT in AdS, obtained by assuming boundary correlation functions consistent with unitarity and a spectrum with the natural two-particle gaps, again as a function of $m_b/m$.
\end{itemize}

\subsubsection{Results}
\begin{figure}
\begin{center}
\includegraphics[width=0.6 \linewidth]{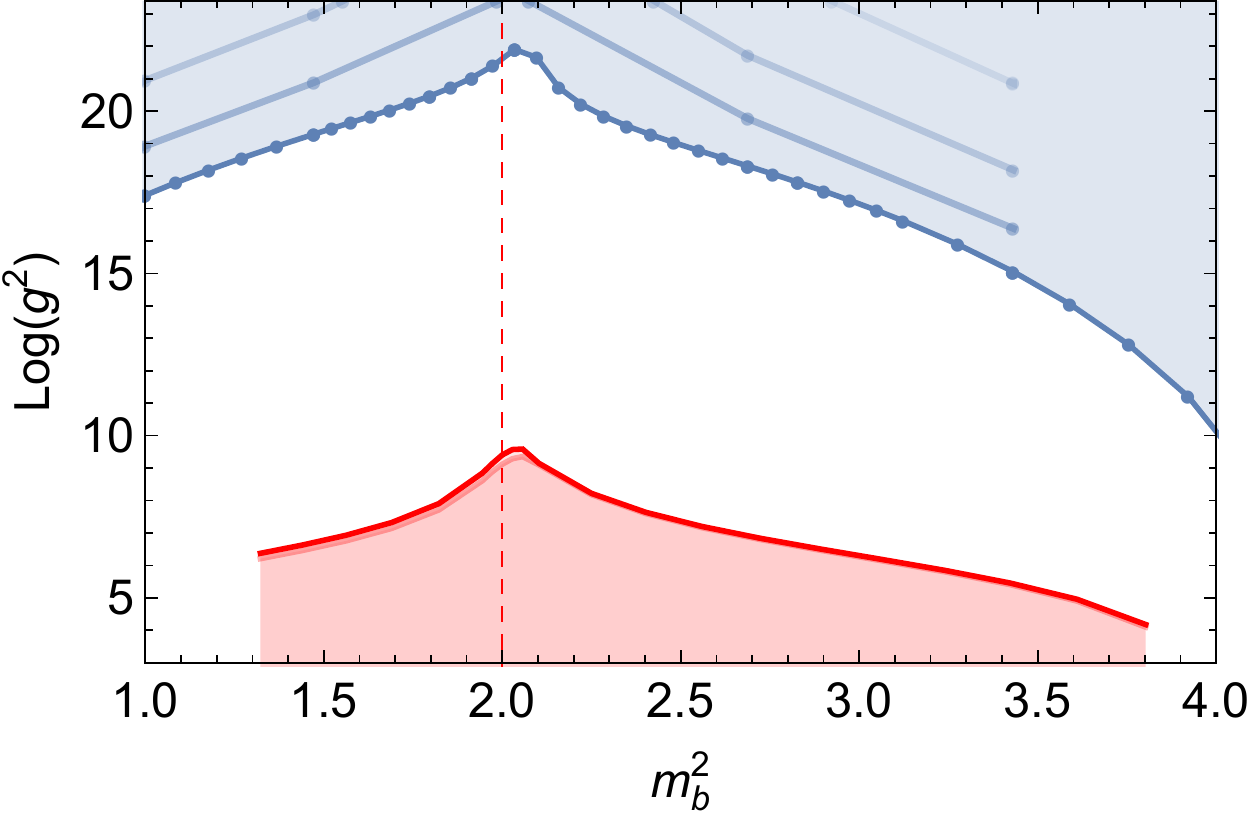}
\caption{\label{fig:bctwithflatspace}Investigating the coupling between two particles of mass $m = 1$ and a third particle of mass $m_b$ in 2+1 dimensional QFT. {\color{red}First approach:} maximum flat-space coupling for a QFT obtained with our ansatz \eqref{eq:Asum}. We plot two curves with $N_{\max}$ equal to 10 (bottom) and 18 (top) which lie almost on top of each other. {\color{mathematicablue}Second approach:} upper bound on the bulk coupling for a QFT in AdS with a radius of curvature $R \approx 16$. The four curves connect the sets of data points which were obtained with functionals with 10 (top), 36, 78 and 136 (bottom) components.}
\end{center}
\end{figure}

The resulting bounds are shown in figure \ref{fig:bctwithflatspace}. Notice the logarithmic scale.\footnote{On a regular scale the shape of the peak is very similar to the one shown in figure \ref{fig:upperbound_residue_4d}.} It is clear that the upper bound obtained from QFT in AdS is way larger than the largest value obtained from the S-matrix bootstrap, but the AdS results have not converged yet and one may hope that the numerical upper bound can decrease much further. The good news, however, is the remarkably similar \textit{shape} of the two curves, both having a somewhat asymmetric peak slightly above $m_b^2 = 2$. In this sense we see a repetition of the results in 1+1 dimensions, namely that we can obtain similar bounds on the residue of a pole in a scattering amplitudes using two drastically different methods.

Physically, it is important to realize that our QFT in AdS approach is completely devoid of any assumptions about the analyticity of the flat-space scattering amplitude. If one agrees that the result in figure \ref{fig:bctwithflatspace} provides evidence of the equivalence between the two approaches, then either our S-matrix bounds on the coupling do not require the amount of analyticity that we have imposed or the analyticity (at least of the extremal scattering amplitudes) is a property that we may hope to derive from the QFT in AdS construction. Either option would be very interesting and should be investigated further.\footnote{In 1+1 dimensions the status of analyticity is a little different. Although we are not aware of any full-fledged two-dimensional proofs, since $t=0$ kinematically one may say that analyticity in two dimensions is similar to forward analyticity in higher dimensions. The analyticity properties of $M(s,t=0)$ can often be proven from axiomatic field theory \cite{Martin:944356}.} 

Although $\Delta = 17$ was the largest value for which we had a full set of results, let us briefly discuss the result for $0 < \Delta \leq 20$. In line with the results in \cite{paperI}, the absolute value of the numerical bounds decreases quickly upon decreasing $\Delta$. For $\Delta \gtrsim 4$ the curve always has a peak hovering around $m_b^2 = 2$, which broadens a bit upon decreasing $\Delta$. For $0 < \Delta \lesssim 4$ the peak moves more or less linearly towards $m_b^2 = 4$ as $\Delta \to 0$. In the future it would be interesting to invest more computational resources and explore in more detail both this behavior and the general convergence of the bounds.

\section{Discussion} \la{secConc}
Here we continued our exploration of the space of S-matrices of gapped quantum field theories initiated in \cite{paperI,paperII}.  We present a fresh approach to an old question of constraining S-matrix elements based on unitarity, crossing and analyticity.  The former two properties are firmly established properties of the S-matrix whose meaning requires no clarification.  By analyticity we mean the rather simplistic (but perhaps most natural) assumption that $M(s,t,u)$ is an analytic function of each of its variables with no singularities in their respective cut planes.  We make no assumption about the properties of the S-matrix outside of this union of cut planes -- i.e. off the physical sheet.

Of course there are many open questions in S-matrix theory pertaining to analyticity.   Are all singularities in the complex Mandelstam variables $s,t,u$ associated to Landau diagrams (as expected based on perturbation theory) or should we be open to more exotic possibilities especially in strongly coupled theories? What is the most general possible large energy behaviour of scattering amplitudes?  Finally, if we bravely cross the gates and delve  into the various Riemann sheets of non-perturbative scattering amplitudes by crossing its various cuts in the physical sheet, what kind of scary Chimeras await us down there? 

We tried to be optimistic -- by assuming the minimal expected singularities in the physical sheet -- and cautious at the same time  -- by assuming as little as possible about the uncontrollable world of the other unphysical sheets or the large energy behaviour of scattering amplitudes. In short we mapped the physical sheet into a few  unit disks and assumed little about the behaviour of amplitudes on the boundary of those disks which is where both the  large energy behaviour as well as the various physical thresholds lie. Inside these disks we assumed that the only singularities were poles associated to stable bound states. 

In the future, it would be interesting to develop new numerical investigations relying on more rigorous analyticity assumptions. Perhaps our results are not too sensitive to this distinction, or perhaps we will encounter exotic S-matrices which make use of the allowed non-analyticity to allow for a wider range of values.  Both would be very interesting!  To this end, it is worth noting that in the case of the quartic pion coupling and the lower bound on the spin zero scattering length we can say with confidence that we are in the former scenario -- our results approach the bounds obtained in \cite{Lopez:1976zs, Auberson:1979ye} and in \cite{Caprini:1980un} which are based on rigorously proven analyticity properties. More evidence for the first scenario is the at least qualitative match between our maximal coupling and the upper bound on the same observable for a QFT in AdS, since the latter computation relied on no analyticity properties whatsoever. Finally we can point to the consistency of our approach with a Mandelstam representation expansion discussed in appendix \ref{Man}.

As for the behaviour at the boundary of the disks the idea here is that we can be agnostic about it and let regular Taylor expansions in the bulk converge towards whatever they \textit{want} to. Of course, without inputing the correct singularities at the boundary of the disk, the numerics should still work but their convergence will suffer considerably. We encountered two examples of this already in the main text. The first is the quartic coupling numerics whose convergence increased substantially once we allowed for a bound state singularity at threshold. Another example is probably the four dimensional bound state coupling numerics when the bound-state mass is less than $\sqrt{2}$ times the mass of the lowest particle. The numerics are converging much slower for that range as clearly seen in the left curves in figure \ref{fig:upperbound_residue_4d}. We suspect in this case it is rather related to a non-trivial large energy behaviour of the S-matrix which the ansatz has a hard time reproducing.\footnote{See appendix \ref{2Dexample} for a two dimensional example which we believe might be the counterpart of what we are observing here.} It would be interesting to investigate this further. 

It is also at the boundary of these disks where we read physical amplitudes with any $s>4m^2$ and negative $t$. Multi-particle production will show up as further cuts at larger $s$ such as $9m^2$, $16m^2$, etc and infinitely many others like $(m+m')^2$, etc if there are other stable particles. 
We saw no signs of these singularities in our numerics. As we for example show in figure \ref{fig:no_particle_production}, our optimal S-matrices do not seem to open multi-particle production cuts in any significant way. A priori this sounds very strange. How could we have no particle production of four particles from two particles if - by crossing one particle to the past - that amplitude is related to a $3\to 3$ process which obviously must exist?\footnote{Of course in $1+1$ there is a well know loophole in this argument which allows for integrable theories \cite{Dorey:1996gd}. This loophole is not possible in higher dimensions.}  Indeed, it is known \cite{Aks} that particle production is mandatory. It can not be strictly zero or it would lead to important contradictions. Unfortunately, the same work \cite{Aks} -- or any other work as far as we know -- does not put a lower bound on \textit{how much} particle production one \textit{must have} and as such we could not reach a sharp contradiction with the numerics which by definition can never rule out an arbitrarily low particle production.\footnote{Actually one can show that a certain amount of production must persist in the limit of infinite spin \cite{Dragt}.  However, to our knowledge, there is no theorem saying, for example, that the first $L$ partial waves exhibit no production.} 

Nonetheless, 
absence of particle production is unphysical in spacetime dimension greater than 2.
We would like to describe more realistic theories where particle production naturally arises. One way of forcing such particle production in a natural way is to study 
multiple S-matrix elements where we consider a system of scattering elements involving not only the lightest particle but also the next-to-lightest etc. We are currently working on this and finding some very encouraging preliminary results in two dimensions where the bounds are often significantly improved and the corresponding S-matrices do exhibit particle production and thus must correspond to genuinely non-integrable theories in contrast to our previous work \cite{paperII}.

The analyticity properties of scattering amplitudes of several particles of different mass are more intricate than what we considered here. The optimistic scenario is that all singularities on the physical sheet follow from Landau diagrams describing propagation of on-shell particles. This \textit{Landau analyticity}  is far from being rigorously established but it is a reasonable physical conjecture to start from. 
Even with this assumption, we will have to deal with anomalous thresholds (singularities that arise from Landau diagrams that are not on a line).
A simple example is the  scattering amplitude of particles of mass greater than $\sqrt{2}$ times the mass lightest particle. 
We plan to analyse this issue in the future, starting in 1+1 dimensions.

\section*{Acknowledgements}

We thank Zoltan Bajnok, Benjamin Basso, Irinel Caprini, Patrick Dorey, Davide Gaiotto, Zohar Komargodski, Martin Kruczenski, Andre Martin, Rafael Porto, Francesco Riva, Slava Rychkov, Amit Sever, Alexander Zamolodchikov and Alexander Zhiboedov for numerous enlightening discussions and suggestions. Research at the Perimeter Institute is supported in part by the Government of Canada through NSERC and by the Province of Ontario through MRI. 
This research received funding from the grant CERN/FIS-NUC/0045/2015.
This work was additionally supported by a grant from the Simons Foundation (JP: \#488649, BvR: \#488659, PV: \#488661)
JP is supported by the National Centre of Competence in Research SwissMAP funded by the
Swiss National Science Foundation.
\newpage
\appendix

\section{\texorpdfstring{$x(s)$ vs $\rho_s,\rho_t$ in $1+1$ dimensions}{x(s) vs rhos, rhot in 1+1 dimensions}} \la{appendixA}
Consider the map $$x(s)=\frac{2-\sqrt{4-s}\sqrt{s}}{s-2}$$ which maps the full $s$-plane minus the cuts $s>4$ and $s<0$ into the unit disc $|x(s)|\le 1$ and the map 
$$\rho_s=\frac{2-\sqrt{4-s}}{2+\sqrt{4-s}} $$
which maps the full $s$-plane minus a single cut $s>4$ into the unit disc $|\rho_s|\le 1$. An analytic function in the $s$-plane minus the cuts $s>4$ and $s<0$ -- such as the S-matrix once we subtract out its known poles -- can be written as 
\beq
f(s)=\sum_{n=0}^\infty c_n \,x(s)^n 
\eeq
Now, we have 
\beq
x(s)=\frac{\rho_s-\rho_t}{1-\rho_s\rho_t}  \qquad \text{where}\qquad  t=4-s
\eeq
which admits a convergent expansion in powers of $\rho_s$ and $\rho_t$ provided they are both inside the unit list (and hence so is their product in the denominator). 
Hence the function $f(s)$ can also be cast as 
\beq
f(s)=\sum_{n=0}^\infty c_{nm} \,\rho_s^n \rho_t^m \qquad \text{where}\qquad  t=4-s
\eeq
As such, our $1+1$ numerics had to work.

\section{Constraint surface in \texorpdfstring{$\rho$}{rho}-coordinates}
\begin{figure}[t]
\begin{center}
\includegraphics[scale=0.6]{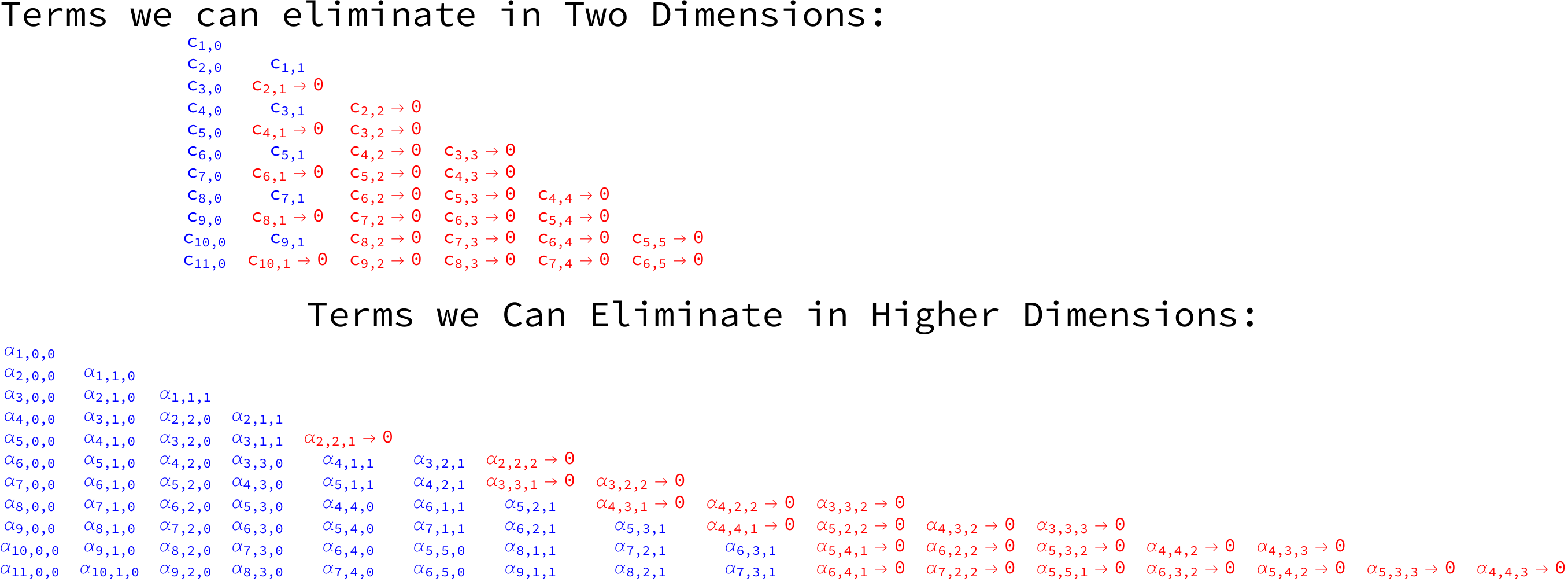} 
\end{center}
\caption{
When centering the $\rho$ variables around general points, we can eliminate all constants $c_{a,b}$ with $a,b>1$ in two dimensions and all constants $\alpha_{a,b,c}$ with $a,b,c>1$ in higher dimensions. By centering the $\rho$ variables around $s_0=2$ in two dimensions and around $s_0=4/3$ in higher dimensions, the kinematical constraints simplify further allowing us to eliminate a few more terms in the Taylor expansions as explained in the text. An option for which terms we can eliminate is illustrated in the tables above up to $N=11$ where the level $N=a+b$ or $N=a+b+c$ is the total powers of $\rho$ in the multiple Taylor expansion. The number of terms we should keep at each level 
is 
$\frac{N}{2}+\frac{(-1)^N}{4}+\frac{3}{4}$ in two dimensions and $\frac{N^2}{12}+\frac{N}{2}+\frac{(-1)^N}{8}+\frac{2}{9} \cos \left(\frac{2 \pi  N}{3}\right)+\frac{47}{72}$ in higher dimensions.}\label{fig:constraints}
\end{figure}

\la{ap:rhoconstraints} 
The on-shell condition imposes
\beq
0=s+t+u-4m^2.
\eeq
If we write this constraint in terms of the $\rho_s, \rho_t$ and $\rho_u$ variables with  arbitrary $s_0$ ({\it cf.} eqn. \ur{rhoVar}) we get a somewhat lengthy expression of the form
\beq
0=\left(s_0-\frac 43 m^2\right) \rho_s^2\,\rho_t^2\,\rho_u^2+\ldots+\left(s_0-\frac 43 m^2\right). 
\eeq
Specializing to the case $s_0=\frac 43 m^2$,  the point $\rho_s=\rho_t=\rho_u=0$ satisfies the on-shell condition. Defining then the symmetrized monomials:
\beq
\rho^{(a,b,c)}=\rho_s^a\,\rho_t^b\, \rho_u^c + \mbox{perms}\ ,
\eeq
the constraint equation becomes ($m=1$):
\begin{eqnarray}
0&=& \rho^{(1,2,2)}-4 \rho^{(1,1,2)}+\rho^{(1,2,0)}+12 \rho^{(1,1,1)}-4 \rho^{(1,1,0)}-\rho^{(1,0,0)}.
\end{eqnarray}
We can now obtain all such constraints by multiplying this equation by other symmetrized monomials. As an example, multiplying by $\rho^{(1,0,0)}$ we get a new identity,
\begin{multline}
0=\rho ^{(0,0,2)}+2 \rho ^{(0,1,1)}-4 \rho ^{(0,1,2)}+\rho ^{(0,1,3)}+2 \rho
   ^{(0,2,2)}-12 \rho ^{(1,1,1)}\\+14 \rho ^{(1,1,2)}-4 \rho ^{(1,1,3)}-8 \rho
   ^{(1,2,2)}+\rho ^{(1,2,3)}+3 \rho ^{(2,2,2)}.
\end{multline}
We can use these identities to systematically reduce the number of monomials in our ansatz as explained in figure \ref{fig:constraints}. Note that in two spacetime dimensions we can set $u=0$ which simplifies the constraint equation to (\ref{constRho}).

\section{Mandelstam Representation}\la{Man}

The double dispersion representation proposed by Mandelstam \cite{Mandelstam:1959bc} implies that the amplitude can be written as follows
\beq
M(s,t,u)=B(s,t)+B(s,u)+B(t,u)\,,
\eeq
where 
\beq
B(s,t)=\int ds' dt'\frac{C(s',t')}{(s'-s)(t'-t)}\, .
\label{eq:doubledispersion}
\eeq
If there are no stable particles below threshold, the double discontinuity $C(s,t)$ has support inside the region   $s>4m^2$ and $t>4m^2$. In practice, this form of the double dispersion relation is not valid and one needs to include subtractions. A simple trick to derive the form of the dispersion relation with $n$ subtractions is to use the identity
\beq
\frac{1}{s'-s}=\frac{(s-s_0)^n}{(s'-s)(s'-s_0)^n}+\sum_{k=0}^{n-1}
\frac{(s-s_0)^{k}}{(s'-s_0)^{k+1}}
\eeq
in equation \eqref{eq:doubledispersion} for both factors in the denominator.
This leads to 
\begin{align}
B(s,t)&=(s-s_0)^n(t-t_0)^n\int ds' dt'\frac{C(s',t')}{(s'-s)(t'-t)(s'-s_0)^n(t'-t_0)^n}
\nonumber  \\
&+\sum_{k=0}^{n-1} 
 (s-s_0)^{k} (t-t_0)^n
 \int  dt'\frac{c_k(t')}{ (t'-t) (t'-t_0)^n} \label{eq:doubledispersionsubtracted}\\
&+\sum_{k=0}^{n-1} 
 (t-t_0)^{k} (s-s_0)^n
 \int  ds'\frac{c_k(s')}{ (s'-s) (s'-s_0)^n}  \nonumber \\
&+\sum_{k,l=0}^{n-1} 
 (s-s_0)^k(t-t_0)^{l}  c_{k,l}  \nonumber
\end{align}
where
\beq
c_k(t)=\int ds'\frac{C(s',t)}{(s'-s_0)^{k+1}}\,,\qquad
c_{k,l}=\int ds'dt'\frac{C(s',t')}{(s'-s_0)^{k+1}(t'-t_0)^{l+1}}\,.
\label{eq:integralsckl}
\eeq
In general the integrals \eqref{eq:integralsckl} do not converge. 
The subtracted dispersion relation is \eqref{eq:doubledispersionsubtracted} considering $c_k(t)$ and $c_{k,l}$ as independent functions from the double discontinuity $C(s,t)$.

Stable particles correspond to delta-function pieces in the single discontinuities $c_k(s)$.\footnote{Therefore, we should use $n\ge j+1$ where $j$ is the maximal spin of the stable particles. In this way, the second and third line of \eqref{eq:doubledispersionsubtracted} can reproduce the degree $j$ polynomial residue of the pole produced by the stable particle.}
Besides these delta-functions, the support of $c_k(s)$ is $s\ge 4m^2$. Therefore, the analytic properties of equation \eqref{eq:doubledispersionsubtracted} imply that 
\beq
B(s,t)={\rm Poles}+\sum_{a,b=0}^\infty \alpha_{(ab)} \rho_s^a \rho_t^b\,,
\eeq
with a convergent double $\rho$ series in the product of two unit disks. This is a more restricted form of formula \eqref{eq:Asum} where we set to zero all coefficients $\alpha_{abc}$ with $a>0$, $b>0$ and $c>0$.

\begin{figure}
\begin{center}
\includegraphics[width=0.6 \linewidth]{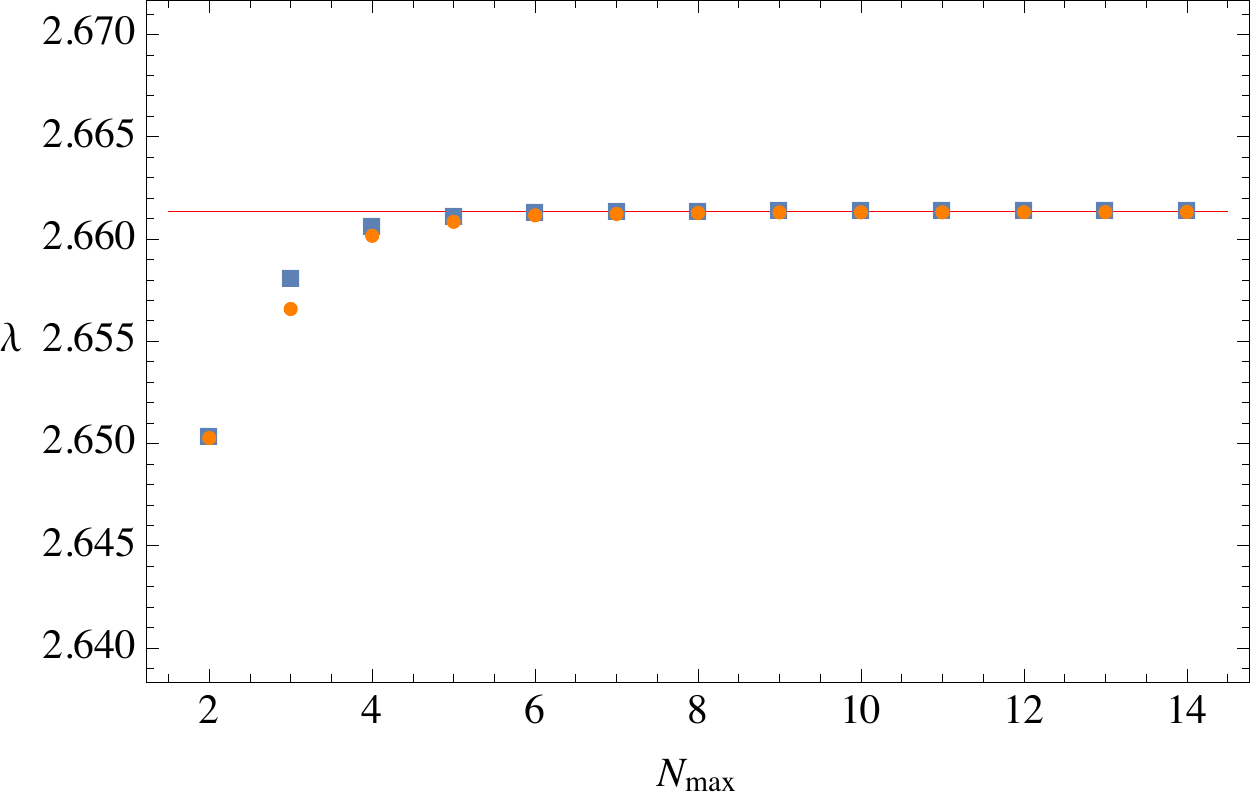}
\caption{\label{pion_mand_comp} Comparison of upper bound on pion coupling using ansatz \eqref{eq:Asum} with $g=0$ and the threshold singularity $\eqref{eq:theshSing}$ included (blue squares) versus ansatz \eqref{mand_sum} (orange dots).  In both cases we use $\ell_{\text{max}}=14$.  The plateau converges to $2.6613...$ in both cases.}
\end{center}
\end{figure}

In order to test the validity of Mandelstam representation, we reconsidered the problem discussed in section \ref{sec:quartic} using the more restricted ansatz 
\beq\label{mand_sum}
B(s,t)= \frac{\alpha}{2} \left(\frac{1}{\rho_s -1} +\frac{1}{\rho_t -1}\right)+\sum_{a,b=0}^{N_{\max}} \alpha_{(ab)} \rho_s^a \rho_t^b\,.
\eeq
In figure \ref{pion_mand_comp}, we show the maximal value of the quartic coupling $\lambda$ obtained with this ansatz. The maximal value $\lambda \approx 2.6613...$ is obtained for $N_{\text{max}} \gtrsim 6$.  This result suggests that in the limit of large $N_{\text{max}}$ both ansatze cover the same space of functions.

\section{Partial Wave Integrals} \la{analyticIntegrals}
\subsection{Pole contributions}
Here we will consider the contribution to partial waves coming from poles of the scattering amplitude. Consider 
\bea
M(s,t,u)_{\mbox{\tiny poles}}=-\frac{g^2}{s-m_b^2}-\frac{g^2}{t-m_b^2}-\frac{g^2}{u-m_b^2}.
\eea
It is easy to compute the partial wave decomposition of this expression. For $d=3$ we get
\beq
\frac{S_\ell(s)-1}{2i}=-\frac{g^2}{32\pi} \frac{\sqrt{s-4}}{\sqrt{s}}\,\left[ \frac{\delta_{l,0}}{s-m_b^2}-\frac{4}{s-4m^2} Q_{\ell}(x_b)\right]\, \label{eq:poled3}
\eeq
with $x_b=x(s,t=m_b^2)$ and $Q_{\ell}(z)$ the Legendre function of the second kind with branch cut along $z\in (-1,1)$. For $d=2$ we instead get
\beq
\frac{S_\ell(s)-1}{2i}=-\frac{g^2}{16\sqrt{s}}\left[\frac{\delta_{\ell,0}}{s-m_b^2}-\frac{2}{m_b \sqrt{s-4m^2+m_b^2}}\left(\frac{m_b-\sqrt{s-4m^2+m_b^2}}{m_b+\sqrt{s-4m^2+m_b^2}}\right)^{\!\!\ell}\  \right].
\eeq

Now consider the contribution to the amplitude from a threshold bound state. The pole part is
\beq
M(s,t,u)_{\mbox{\tiny th.pole}}=-\frac{2\alpha}{\sqrt{6}\sqrt{4m^2-s}}-\frac{2\alpha}{\sqrt{6}\sqrt{4m^2-t}}-\frac{2\alpha}{\sqrt{6}\sqrt{4m^2-u}}.
\eeq
If we focus on the case $d=3$, we must compute integrals of the form:
\beq
\int_{-1}^1 \ud x\frac{P_{\ell}(x)}{\sqrt{4m^2-t(x)}} 
\eeq
with $t(x)=-\frac 12 (s-4)(1-x)$. Introducing the generating function for the Legendre polynomials
\bea
\sum_{n=0}^{+\infty} z^n P_n(x)=\frac{1}{\sqrt{1-2 x z+z^2}},
\eea
it is not difficult to obtain
\bea
\int_{-1}^1 \ud x \frac{ P_{\ell}(x)}{\sqrt{4m^2-t(x)}} =\frac{4}{2\ell+1}\,\frac{(\sqrt{s}-2m)^{\ell}}{(\sqrt{s}+2m)^{\ell+1}}
\eea
Adding up contributions from $s,t,u$  the partial amplitudes are
\bea
\frac{S_\ell(s)-1}{2i}=-\frac{\alpha}{16\sqrt{6}\pi}\frac{\sqrt{s-4m^2}}{\sqrt{s}}\left(\frac{i\,\delta_{\ell,0}}{\sqrt{s-4m^2}}+\,\frac{4}{2\ell+1}\,\frac{(\sqrt{s}-2m)^{\ell}}{(\sqrt{s}+2m)^{\ell+1}}\right)\label{eq:thpolewaves}
\eea

\subsection{\texorpdfstring{$\rho_s\, \rho_t\, \rho_u$}{rhos, rhot, rhou} contributions}
Here we will show how to obtain the contribution to the partial amplitudes from terms of the form $\rho_s^a\,\rho_t^b\, \rho_u^c$ analytically in $d=3$. While the calculation is somewhat tedious, the underlying concept is simple:  the integral that we want to do has only one cut (of square-root type) in the integrand and thus with a simple trigonometric change of variables the integrand can be converted to a rational function and computed by partial fractions (or some more clever method).  

The non-trivial integrals to perform take the form
\bea
I^{\ell}_{\ b,c}=\int_{-1}^1 \ud x P_{\ell}(x)\, \rho(t)^b \rho(u)^c
\eea
with, as in \eqref{rhoVar} with $m = 1$,
\bea
\rho(s)=\frac{1-\sqrt{1-\frac{s-s_0}{4-s_0}}}{1+\sqrt{1-\frac{s-s_0}{4-s_0}}}
\eea
In applications we typically set $s_0=4/3$. We next introduce our first inspired change of variables from $x$ to $\phi$ which is given by
\bea
x=-\frac{s+4}{s-4}\, \cos(2 \phi).
\eea
In these variables we get:
\bea
I^{\ell}_{\ b,c}=4\,\left(\frac{s+4}{s-4}\right)\,\, \int \ud \phi \,P_{\ell}(x(\phi)) \sin(\phi)\cos(\phi) \left(\frac{1-r \cos \phi}{1+r \cos \phi}\right)^b\,\left(\frac{1-r \sin \phi}{1+r \sin \phi}\right)^c
\eea
where we also introduced
\beq
r^2 \equiv \frac{4+s}{4-s_0}\,.
\eeq
We should now do the usual change of variables,
\bea
\phi=2\,\arctan(y)
\eea
This gives
\bea
I^{\ell}_{\ b,c}= \frac{4^{2}r^2}{r^2-\frac{8}{4-s_0}} \, \int \ud y \,P_{\ell}(x(y))\,\frac{y(1-y^2)}{(1+y^2)^3}\, \left(\frac{(1-r)+y^2(1+r)}{(1+r)+y^2(1-r)}\right)^b\,\left(\frac{1-2r\,y+y^2}{1+2 r\,y+y^2}\right)^c.
\eea
We have
\bea
x(y)=\frac{r^2}{r^2-\frac{8}{4-s_0}}\, \frac{1-6 y^2+y^4}{(1+y^2)^2},
\eea
and the integral runs from $y_i$ to $y_f$ with
\bea
y_i=\frac{\sqrt{4-s_0}}2\, \left(r-\sqrt{r^2-\frac{4}{4-s_0}}\right), \qquad y_f=\left(\frac{r-\frac{2}{\sqrt{4-s_0}}}{r+\frac{2}{\sqrt{4-s_0}}}\right)^{\frac 12}.
\eea
The trick now is to rewrite the integration region using the discontinuity of a logarithm,
\bea
\int_{y_1}^{y_2} \ud y f(y)=\frac{1}{2 \pi i} \int_{y_1}^{y_2}
\ud yf(y)\,\mbox{Disc} \log\left(\frac{y-y_2}{y-y_1}\right)=\frac{1}{2 \pi i} \int_{(y_1,y_2)} \ud yf(y)\, \log\left(\frac{y-y_2}{y-y_1}\right)
\eea
where $(y_1,y_2)$ is a clockwise contour wrapping the line segment from $y_1$ to $y_2$. In our case $f(y)$ is a rational function, therefore we can  pull the contour to infinity so that it picks up the poles of $f(y)$  to obtain exact expressions.

\subsection{Large energy}
\label{subapp:largeenergy}
Let us consider the large energy limit $s\to \infty$ of our ansatz. Since unitarity is imposed for each spin $\ell$, we are interested in the limit $s\to \infty$ with fixed scattering angle $\theta$. In this limit, we find
\begin{align}
\rho_s^a \rho_t^b \rho_u^c&= (-1)^{a+b+c}
\left[1+
\frac{2\sqrt{4-s_0}}{\sqrt{s}}
\left(ia-\frac{\sqrt{2}b}{\sqrt{1-x}}-\frac{\sqrt{2}c}{\sqrt{1+x}}\right)
\right.
\nonumber \\
&\left.-\frac{4\sqrt{2}(4-s_0)a}{s}
\left(\frac{ib}{\sqrt{1-x}}+\frac{ic}{\sqrt{1+x}}+{\rm real}\right)
+O\left( s^{-\frac{3}{2}}\right)\right]\,.
\label{eq:rhoabclarges}
\end{align}
The contribution from the pole terms in our ansatz are real and of order $1/s$ in this limit and therefore can be neglected.
The leading term in \eqref{eq:rhoabclarges} only contributes to the spin 0 partial wave. The large $s$  expansion of $S_0(s)$ is given by
\bea
S_0(s)=1+\frac{i s^{\frac{d-3}{2}}}{2^{2d-1}\pi^{\frac{d}{2}-1}
\Gamma\left(\frac{d}{2}\right)} \sum_{a,b,c} \alpha_{abc}(-1)^{a+b+c}
\left[1+\frac{2ia\sqrt{4-s_0}+{\rm real}}{\sqrt{s}}+O\left(\frac{1}{s}\right)
\right]\,.
\eea
Unitarity implies that (for $d>4$ the inequality must be saturated)
\bea
\sum_{a,b,c} \alpha_{abc}(-1)^{a+b+c}a\ge 0\,.
\eea
If $d>2$ then unitarity also implies that
\bea
\sum_{a,b,c} \alpha_{abc}(-1)^{a+b+c}=0\,.
\eea
For $d=2$, the correct condition is
\bea
32 \sqrt{4-s_0} \sum_{a,b,c} \alpha_{abc}(-1)^{a+b+c}a  \ge
\left[ \sum_{a,b,c} \alpha_{abc}(-1)^{a+b+c}\right]^2 \,.
\eea

For $\ell>0$ (even) we find
\bea
S_\ell(s)=1-i s^{\frac{d-4}{2}} I_\ell 
\sum_{a,b,c} \alpha_{abc}(-1)^{a+b+c}a \left[ 1 
+\frac{i2\sqrt{4-s_0}b+{\rm real}}{\sqrt{s}} 
+O\left( \frac{1}{s}\right) \right]
\,,
\eea
where \footnote{
For $d\le 2$ the integral $I_\ell$ is divergent. The origin of this divergence is that the we can only use the large $s$ form of the integrand for $(1-x) s\gg 1$ and $(1+x) s \gg 1$. The effect of this can be taken into account by including the $s$-dependence $I_\ell \sim s^{\frac{2-d}{2}}$ for $d<2$ and 
 $I_\ell \sim\log s$ for $d=2$.}
\bea
I_\ell= 4\sqrt{2}\sqrt{4-s_0}\int_{-1}^1 dx (1-x^2)^\frac{d-3}{2} P_\ell^{(d)}(x) \frac{1}{\sqrt{1+x}}>0\,.
\eea
Therefore, unitarity implies (for $d>5$ the inequality must be saturated)
\bea
\sum_{a,b,c} \alpha_{abc}(-1)^{a+b+c}ab\le 0\,.
\eea 
For $d>3$ unitarity also implies 
\bea
\sum_{a,b,c} \alpha_{abc}(-1)^{a+b+c}a= 0\,.
\eea
For $d=3$ we find
\bea
40\pi \sum_{a,b,c} \alpha_{abc}(-1)^{a+b+c}ab  \le-
\left[ \sum_{a,b,c} \alpha_{abc}(-1)^{a+b+c}a\right]^2 \,.
\eea
where we used that  $I_{\ell} < I_2 =\frac{\sqrt{4-s_0}}{10\pi}$ for $\ell>2$.

Where applicable, we have verified the above constraints a posteriori for our numerical solutions and found them satisfied to very good numerical accuracy.

As a final comment, we remark that the unitarity constraints dictate that $\lim_{s\to \infty} S_\ell(s)=1$ for any amplitude within our ansatz with finite $N_{\max}$.\footnote{With the exception of  $d=5$  where it is possible to obtain $\lim_{s\to \infty} S_\ell(s)\neq 1$ for $\ell>0$.}
This property is likely to be too restrictive, and it is therefore worthwhile to try to improve our ansatz with more singular terms compatible with unitarity and analyticity. As a first attempt we added an extra term of the form $(\rho_s + 1)(\rho_t + 1)^{-1}$ plus $s$, $t$, $u$ permutations, which allows $\lim_{s \to \infty} (S_\ell(s) - 1)$ to be non-zero -- this modification however did not significantly change any of the results displayed above. In the future we plan to add other more singular terms and investigate their effect in more detail.\footnote{We also deem it likely that there exists a higher-dimensional version of the two-dimensional construction discussed in footnote \ref{essentialsingularitycounterexamplefootnote} that would lead to unbounded couplings, but we again expect the associated essential singularity to be in conflict with causality.} Finally, the restricted behavior at large $s$ might also be a source of slow convergence when $N_{\rm max} \to \infty$ we have observed in some cases. This idea is also corroborated by the two dimensional analysis in appendix \ref{2Dexample}.

\subsection{Large spin}

The partial waves can also be written in terms of an hypergeometric function,
\beq
P^{(d)}_\ell(x)= 
\frac{2^{1-2 d} \pi ^{\frac{1}{2}-\frac{d}{2}} }{\Gamma \left(\frac{d-1}{2}\right)}\,
   _2F_1\left(-\ell,d+\ell-2;\frac{d-1}{2};\frac{1-x}{2}\right)\,.
\eeq
It is convenient to define
\beq
Q^{(d)}_\ell(x)= -
\frac{ \Gamma (l+1)
   (x-1)^{2-d-\ell} }{
   \pi ^{\frac{d}{2}-1} 2^{2 d+\ell-1}\Gamma
   \left(\frac{d}{2}+\ell\right)}
   \, _2F_1\left(d+\ell-2,
   \frac{d+2\ell-1}{2} ;d+2 \ell-1;\frac{2}{1-x}\right)
\eeq
such that
\beq
{\rm Disc} \left[(x^2-1)^{\frac{d-3}{2}}Q^{(d)}_\ell(x) \right]=
2\pi i (1-x^2)^{\frac{d-3}{2}}P^{(d)}_\ell(x)\,, \qquad -1<x<1\,.
\eeq
Notice that for integer $d$ the function $Q^{(d)}_\ell(x)$ has no monodromy around $x=\infty$. We will work in the sheet where $Q^{(d)}_\ell(x)$ only has a branch cut from $x=-1$ to $x=1$.
The factor $(x^2-1)^{\frac{d-3}{2}}=x^{d-3} (1-x^{-2})^{\frac{d-3}{2}} $ has the same analytic properties.
Then we can write 
\beq 
  \int\limits_{-1}^1\! dx\, (1-x^2)^{\frac{d-3}{2}}P^{(d)}_\ell(x)  M(s, x) = \frac{1}{2\pi i} \oint_C dx (x^2-1)^{\frac{d-3}{2}}Q^{(d)}_\ell(x)
  M(s,x)\,,
\eeq
where the contour $C$ encircles the real segment $[-1,1]$ clockwise and $M(s,x)$ denotes the amplitude $\left. M(s, t)\right|_{t\to \tfrac{1}{2}(s-4) (x-1)}$.
Since
\beq
Q^{(d)}_\ell(x)\approx -\frac{1}{
2^{\frac{3d}{2}  } \pi
   ^{\frac{d-2}{2} }
   \ell^{\frac{d-2}{2}}
   \left(x^2-1\right)^{\frac{d-2}{4}}} 
   \frac{1}{(x+\sqrt{x^2-1})^{\ell+\frac{d-2}{2}}}\,,
\eeq
for large $\ell$ and $x^2 >1$, we can expand the contour
and drop the contribution from infinity. At large spin, the integral will be dominated by the singularity of $M(s,x)$ closer to the origin $x=0$.
Generically, this will come from the poles associated with stable particles. More precisely,
\begin{align}
&\int\limits_{-1}^1\! dx\, (1-x^2)^{\frac{d-3}{2}}P^{(d)}_\ell(x)  M(s, x) = \frac{1}{2\pi i} \oint_C dx (x^2-1)^{\frac{d-3}{2}}Q^{(d)}_\ell(x)
  M(s,x) \\
&= -\frac{1}{2\pi i  }\left[
 \int_{-\infty}^{-x_1(s)}  dx + \int_{x_1(s)}^\infty dx \right]
 (x^2-1)^{\frac{d-3}{2}}Q^{(d)}_\ell(x) \left[M(s,x+i\epsilon)-M(s,x-i\epsilon)\right]\\
  &= -\frac{1}{i\pi  }  \int_{x_1(s)}^\infty dx 
 (x^2-1)^{\frac{d-3}{2}}Q^{(d)}_\ell(x) \left[M(s,x+i\epsilon)-M(s,x-i\epsilon)\right]
\end{align}
where $x_1(s)$ is determined from
\beq
t(s,x)=m_1^2\qquad \Rightarrow \qquad
x_1(s)=1+\frac{2m_1^2}{s-4m^2}\,.
\eeq
In fact, the pole $\frac{-g^2}{t-m_1^2}$ contributes
\beq
\int\limits_{-1}^1\! dx\, (1-x^2)^{\frac{d-3}{2}}P^{(d)}_\ell(x)  M(s, x) \approx -\frac{2g^2}{s-4m^2}(x_1(s)^2-1)^{\frac{d-3}{2}}Q^{(d)}_\ell(x_1(s))\,,
\label{polecontribution}
\eeq 
which decays exponentially with $l$. Notice that this gives a purely imaginary contribution to $S_\ell(s)$ (see equation \eqref{eq:partialwaves}), which by itself would violate unitarity. However, unitarity can be restored with a small real contribution of the order of the square of \eqref{polecontribution}.
At large $l$, this requires that we match the exponential behaviour
\beq
 \left(x_1(s)+\sqrt{x_1(s)^2-1}\right)^2 = x_2(s)+\sqrt{x_2(s)^2-1}  \qquad
\Leftrightarrow \qquad \frac{m_2^2}{4m_1^2} = 1+\frac{ m_1^2}{s-4m^2} \,.
\eeq
In other words, unitarity can be restored with another particle or particles of total invariant mass squared   $m_2^2\ge 4m_1^2$. 
This is what happens in perturbation theory where the box diagram restores unitarity of the tree-level exchanges.

Let us now study the contribution from the polynomial terms $\rho_s^a\rho_t^b\rho_u^c$ in our ansatz. The discontinuity of $M$ for $x>1$ comes from
\beq
\rho(t(s,x+i\epsilon))^{b} -\rho(t(s,x-i\epsilon))^{b} \approx
2i b \sqrt{\frac{2s-8m^2}{4m^2-s_0}} \sqrt{x-x_\star(s)}
\eeq
where 
\beq
x_\star(s)=\frac{s+4m^2}{s-4m^2}
\eeq
and we only kept the leading behaviour of the discontinuity near its lower end $x_\star(s)$.
Similarly, we can approximate
\beq
Q^{(d)}_\ell(x)\approx Q^{(d)}_\ell(x_\star(s)) \exp\left[{-\ell\frac{s-4m^2}{4m\sqrt{s}}(x-x_\star(s))}\right]
\eeq
and find
\begin{align}
&\int\limits_{-1}^1\! dx\, (1-x^2)^{\frac{d-3}{2}}P^{(d)}_\ell(x) \rho_s^a\rho_t^b\rho_u^c 
\approx
-\frac{1}{i\pi  } \rho_s^a \rho_{-s}^c 
(x_\star(s)^2-1)^{\frac{d-3}{2}}Q^{(d)}_\ell(x_\star(s)) \times \\
&\times \int_{x_\star(s)}^\infty dx 
\exp\left[{-\ell\frac{s-4m^2}{4m\sqrt{s}}(x-x_\star(s))}\right]
  2i b \sqrt{\frac{2s-8m^2}{4m^2-s_0}} \sqrt{x-x_\star(s)}
  \\
  =&-\frac{b}{ \sqrt{\pi} \ell^{\frac{3}{2} }}
   \frac{8m^{\frac{3}{2} }s^{\frac{3}{4} }}{s-4m^2} 
    \sqrt{\frac{2 }{4m^2-s_0}} 
   \rho_s^a \rho_{-s}^c 
(x_\star(s)^2-1)^{\frac{d-3}{2}}Q^{(d)}_\ell(x_\star(s)) \\
\approx&  \frac{F(s)}{\ell^\frac{d+1}{2}  \left(x_\star(s)+\sqrt{x_\star^2(s)-1}\right)^\ell}   \,b\rho_s^a \rho_{-s}^c
\end{align}
where $F(s)>0$ for $s>4m^2$. Notice that at large $\ell$ the leading contribution comes from $t \approx 4m^2$ which implies that $\rho_u \to \rho_{-s}$. Unitary implies ${\rm Re}\, S_\ell(s)\le 1$  which at large $\ell$ becomes
\beq
\sum_{a,b,c}\alpha_{abc} \,b\left({\rm Im}\,\rho_s^a\right) \rho_{-s}^c \ge 0\,.
\label{eq:largelunitarity}
\eeq
Notice that this condition is independent of the spin $\ell$ and of the spacetime dimension $d$. This justifies our numerical procedure of truncating the unitarity conditions at some value of the spin $\ell_{\rm max}\gg 1$.
Writing $\rho_s=e^{i\phi}$ with $\phi\in [0,\pi]$, equation (\ref{eq:largelunitarity}) can be written as
\beq
\sum_{a,b,c}\alpha_{abc} \,(-1)^c  b \sin( a \phi) 
\left[
\frac{\sqrt{1+y_0 \cos\frac{\phi}{2}} - \cos\frac{\phi}{2}}
{\sqrt{1+y_0 \cos\frac{\phi}{2}} + \cos\frac{\phi}{2}}
\right]^c \ge 0\,,\qquad \forall \,\phi\in [0,\pi]\,,
\eeq
where $y_0=\frac{4m^2+s_0}{4m^2-s_0}>-1$.

The constraints \eqref{eq:largelunitarity} are linear constraints on the numerical coefficients and can easily be taken into account in our numerical code (again by sampling for a discrete set of values of $s$). We have run several of our analyses both with and without this additional constraint. As expected, the effect of the additional term decreases with the maximum spin $\ell_{\max}$ for which we manifestly check the unitarity constraints. For the values $\ell_{\max}$ used in our plots the effect of including \eqref{eq:largelunitarity} is always small and amounts to maybe to a one percent change in the final result.

\subsection{Threshold Expansion and Elastic Unitarity}
\label{app:thresholds}
Here we shall discuss the threshold behaviour of amplitudes satisfying our ansatz. We start with the expression for the amplitude,
\bea
M(s,t,u)=\sum_{a,b,c=0}^{+\infty}\alpha_{a b c}\, \rho_s^a \rho_t^b \rho_u^c +\mbox{poles}
\eea
At threshold the poles become constants and are irrelevant. This is not so for threshold poles which are discussed separately below. Define $w:=\sqrt{s-4}$.  Then for $s\to 4^+$ above the cut we have
\begin{subequations}
\bea 
\rho_s&=&1+2\sum_{n=1}^{+\infty} \left(\frac{i}{\sqrt{4-s_0}}\right)^n w^n=1+\frac{2 i}{\sqrt{4-s_0}} w+\ldots\\
\rho_t^b \rho_u^c
&=& \sum_{k=0}^{\infty}w^{2k} \left(\sum_{n+m=k} c_{n,m}  (1-x)^n (1+x)^m\right)
\eea
\end{subequations}
Recall that in our conventions the partial waves take the form:
\bea
S_{\ell}(s)=1+i \frac{(s-4)^\frac{d-2}{2}}{ \sqrt{s}}   \int\limits_{-1}^1\! dx\, (1-x^2)^{\frac{d-3}{2}}P^{(d)}_\ell(x) \left. M(s, t)\right|_{t\to \tfrac{1}{2}(s-4) (x-1)}.
\eea
The leading contribution for the spin $\ell$ partial wave corresponds to the $k=\ell$ term in the above, leading to
\bea
S_\ell(s)&=1-b_\ell w^{d-1+2\ell}+i a_\ell w^{d-2+2\ell} +\ldots,  
\eea
with real $a_{\ell},b_\ell$. These are linear combinations of the coefficients $\alpha_{abc}$ in our ansatz. Unitarity near threshold imposes: 
\begin{subequations}
\begin{align}
b_\ell&\geq 0, &d\geq 2,\ell\geq 0; \\
a_0&=0, &d=2, \ell=0;\\
b_0&\geq a_0^2/2, &d=3, \ell=0.
\end{align}
\end{subequations}

Near threshold we have the expansion of $S_{\ell}(s)$ in terms of the phase shift,
\bea
S_{\ell}(s)=e^{2 i \delta_\ell(s)}\sim 1+2i \delta_{\ell}(s)-2 \delta_{\ell}(s)^2+\ldots.
\eea
Absence of particle production would imply reality of $\delta_{\ell}(s)$, and hence a measure of the inelasticity of the amplitude at the threshold is
\bea
\frac{\mbox{Re}\, [1-S_l(s)]}{\left[\mbox{Im}\, S_\ell(s)\right]^2} =  O[(s-4)^{-\ell}].
\eea
We see that for positive spin we generically get a divergent result in the threshold limit. This means that our ansatz does not automatically give an amplitude which becomes purely elastic as we approach threshold, unlike what we would expect on physical grounds. In order for purely elastic scattering to hold, we would have had to impose order $\ell$ linear constraints on the coefficients of the threshold expansion of the spin $\ell$ partial wave. We did {\it not} impose these in our numerical computations. However, experimentally we do find that as the number of parameters in our ansatz is increased, the coefficients in the threshold expansion seem to decrease.

\section{Non-Relativistic Limit}\la{NR}
Consider a scalar $\phi$ of mass $m$ interacting with itself via the exchange of a second heavy scalar $\Phi$ with mass $m_{b}=2m-\epsilon$ with small $\epsilon$.  We can think of $\Phi$ as a loosely bound state of two $\phi$ particles with binding energy $\epsilon$. The two body amplitude for $\phi+\phi$ scattering contains a pole at $s=m_{b}^{2}$ due to virtual production of a $\Phi$ which is just below the the two-particle threshold at $s=(2m)^{2}$.  The residue of this pole $g^{2}$ is the square of the  $\phi\phi\Phi$ coupling. Now consider low energy $\phi+\phi$ scattering and write $s=(2m+E)^{2}$ where $E$ is the centre of mass energy after subtraction of the rest mass.  The $s$-channel pole of the amplitude is given by\footnote{The factor $m^{5-d}$ is to make the coupling $g^2$ dimensionless.}
\beq
M^{\text{pole}}=\frac{m^{5-d}g^{2}}{s-m_{b}^{2}}\sim \frac{m^{5-d}g^{2}/\e}{4m(E/\e+1)}\label{low_E_T_pole}
\eeq
where we have assumed small $E$ and $\e$.   The $l=0$ phase shift inherits this pole through the relation
\beq\label{phase_shift}
\, \frac{\sqrt{s}}{i (s-4m^{2})^{\frac{d-2}{2}}}(S_{0}(s)-1) =\int\limits_{-1}^1\! dx\, (1-x^2)^{\frac{d-3}{2}}P^{(d)}_0(x) \left. M(s, t)\right|_{t\to \tfrac{1}{2}(s-4) (x-1)}
\eeq
Plugging \eqref{low_E_T_pole} into \eqref{phase_shift} and zooming in on the pole of the phase shift at $s=(2m-\e)^{2}$ we have
\beq\label{phase_shift2}
2^{3-d}m (m \e)^{1-d/2} S^{\text{pole}}_{0}(E/\e) \sim \frac{2^{1-2 d} \pi ^{1-\frac{d}{2}}}{\Gamma \left(\frac{d}{2}\right)}  \frac{m^{5-d}g^{2}/\e}{4m(E/\e+1)}
\eeq
We write the pole of the phase shift as $g_{NR}^{2}/(E/\e+1)$ where $g_{NR}^{2}$ is the residue in units of the binding energy $\e$.  We then have
\beq\label{nonrel_g}
g^{2} \to \, 2^{4+d} \pi^{\frac{d}{2}-1}\Gamma(d/2) g_{NR}^{2} (\e/m)^{2-\frac{d}{2}} 
\eeq
We will show below that there is a bound on the non-relativistic coupling $g_{NR}^{2}\le 2^{2}$.  Note that this correctly predicts the behaviour
\beq
 g^{2}_{1+1}\le 2^{7} (\e/m)^{3/2} 
\eeq 
in $1+1$ dimensions \cite{paperII}.   Moreover, this limit has been studied extensively in $3+1$ dimensions ($d=3$) \cite{Ruderman1958,Gribov1961}.  These authors find (adding a factor of 2 to their results to account for identical particles)
\beq
g_{3+1}^{2}\le 2^{8} \pi \sqrt{\e/m}
\eeq 
and thus we find perfect agreement with \eqref{nonrel_g}.

Let us now derive the bound on $g_{NR}^{2}$ quoted above.  Recall that we are considering a very weakly bound state with binding energy $\e$.  We wish to obtain the behaviour of $g_{\text{max}}^{2}(\e/m)$ for small $\e/m$.  Thus we concentrate on ``slow'' physics at energies $E\sim\e$ (recall $E$ is the centre of mass energy after removal of the rest mass).  Formally, in the phase shift we consider $s\to \bar{s}\e^{2}$ and consider finite $\bar{s}$ as $\e \to 0$.  Any singularites of the phase shift that are a finite distance (in $s$) from the two-particle threshold -- e.g. the left cut and inelastic thresholds -- will be infinitely far away in $\bar{s}$ and thus only contribute through positive powers of $\e$.  We can thus neglect these singularities to obtain the leading behaviour of $g_{\text{max}}^{2}(\e/m)$ and consider a non-relativistic phase shift $S_{\text{NR}}(\bar{E})$ with only a right-hand cut starting at $\bar{E}=0$ and a single bound-state pole at $\bar{E}=-1$, where $\bar{E}=E/\e$.   Since this phase shift is bounded by unitarity along the cut and cannot grow faster than a constant at infinity then the residue of the pole can easily be bounded by maximum modulus theorem.  Perhaps the cleanest way to derive the precise value of the bound is to consider the change of coordinates
\beq
x(E)=\frac{1-(-\bar{E})^{1/2}}{1+(-\bar{E})^{1/2}},\;\;\;\;\;  \bar{E}(x)=-\frac{(x-1)^{2}}{(x+1)^{2}}
\eeq
which maps the E-plane minus the positive real axis to the unit disk and maps the bound state pole to the origin 
\beq
\frac{g_{NR}^{2}}{\bar{E}(x)-1} \sim \frac{g^{2}_{NR}}{4\, x}
\eeq
Now note that the function $f(x)=x \,S_{\text{NR}}(x)$ is analytic throughout the unit disk and obeys $|f|\le 1$ on the boundary due to unitarity.  Thus maximum modulus theorem implies $1\ge f(0) = g_{NR}^{2}/4$ which is the desired bound.

\section{Semidefinite programming implementation} \label{app:numerics}

Consider an ansatz as in \eqref{eq:Asum}, truncated such that $a+b+c \leq N_{\max}$. After eliminating the redundant monomials as described in appendix \ref{ap:rhoconstraints}, we are left with a finite subset of the $\alpha_{abc}$, which together with the coupling $g^2$ completely determine the amplitude. Let us group these real coefficients into a vector that we call $\vec \eta$, so we can schematically write \eqref{eq:Asum} as
\beq
M(s,t,4-s-t) = \vec{\eta} \cdot \overrightarrow{M(s,t)}
\eeq
with $\overrightarrow{M(s,t)}$ the vector of functions of $s$ and $t$ that each coefficient multiplies. We then substitute into the partial amplitude projection \eqref{eq:partialwaves} and get, schematically,
\beq
S_\ell(s) = 1 + i \vec{\eta} \cdot \overrightarrow{f_\ell(s)}
\eeq
with $\overrightarrow{f_\ell(s)}$ defined in the obvious way as the integral of $\overrightarrow{M(s,t)}$ against the Gegenbauer polynomials with the right prefactor. The unitarity constraints $|S_\ell(s)|^2 \leq 1$ now dictate that for all physical $\ell$ and $s$ we must have
\beq
\left(1 - \vec{\eta} \cdot \vec I\right)^2 +( \vec\eta \cdot \vec R)^2 \leq 1 \qquad \Leftrightarrow \qquad U \equiv 2 \vec \eta \cdot \vec I - (\vec \eta \cdot \vec I)^2- (\vec\eta \cdot \vec R)^2 \geq 0
\eeq
with $\vec{R} = \text{Re}[\overrightarrow{f_\ell(s)}]$ and $\vec{I} = \text{Im}[\overrightarrow{f_\ell(s)}]$. This constraint can be re-phrased as a semidefiniteness condition. Indeed, consider the matrix
\bea
M:=\left ( \begin{array}{cc} 1+ \vec \eta \cdot \vec R & 1-\vec{\eta} \cdot{\vec I}\\ 1-\vec{\eta} \cdot{\vec I} & 1-\vec \eta \cdot \vec R \end{array}\right)
\eea
The eigenvalues of this matrix are precisely
\bea
\lambda_{\pm}=1\pm \sqrt{1 - U}
\eea
As befits a Hermitian matrix, they are always real since $U \leq 1$ by construction. It is now clear that
\bea
M\succeq 0 \Leftrightarrow U \geq 0.
\eea
and the unitarity constraints are therefore precisely those of a semidefinite program.

We need to choose a grid of values of $s$ and a finite set of spins $\ell$ for which to test the unitarity constraints. We found it sufficient to take approximately 200 values of $s$, interspersed uniformly along the upper half of the unit circle in the $\rho_s$ variable defined in the main text. We observed no significant change in the results by taking a more refined $s$ grid, or by distributing the points differently along the unit circle. The maximal value of the spin $\ell_{\max}$ is indicated in the various plots. Notice that $\ell_{\max}$ needs to be sufficiently big since otherwise the extremal value completely destabilizes -- see for example the data points in figure \ref{pion_upper_bound1} with $\ell_{\max} = 10$ for large $N_{\max}$. In practice we observed convergence by taking $\ell_{\max}$ at least as large as $N_{\max}$, and for the scattering length computations we needed at least $N_{\max} + 4$. Increasing $\ell_{\max}$ beyond these values did not affect our results.

In our numerical computations we did find it necessary to retain very high precision, generally at least 1000 binary digits. This appears to stem from the approximate redundancy that remains even after imposing the polynomial constraint \ref{ap:rhoconstraints}. To illustrate this we can for example compute a derivative like
\beq
\begin{split}
&\frac{\partial^2}{\partial s^2} \left.\left( \sum_{a,b,c} \alpha_{abc} \rho^a(s) \rho^b(t) \rho^c(4 - s - t) \right)\right|_{s = t = 4/3} = \\
& \qquad \frac{9}{256}\left( \a_{100} + \a_{001} + \frac{1}{2} \a_{200} + \frac{1}{2} \a_{002} - \frac{1}{2} \a_{101} \right)
\end{split}
\eeq
In a typical solution we find that this derivative is rather modest in magnitude, of order $10^2$ or so, whereas the individual coefficients can be very large, of order $10^{24}$ in some solutions. These kind of cancellations require high precision.

We have performed all the numerical computations in section \ref{sec4} with \texttt{sdpb} \cite{Simmons-Duffin:2015qma}. Details of the computations like parameter settings are available from the authors upon request.

\section{Slow convergence on a simple 2D example} \la{2Dexample}
In this appendix we revisit once more the two dimensional problem considered in section \ref{sec2} but this time done in the language of the $M$ amplitude rather than $S$. In two dimensions the two are simply related by 
\beq
S(s,t)-1 =\frac{1}{2\sqrt{s t}} \times M(s,t) \,, \qquad s+t=4m^2 \,.  \la{STrelation}
\eeq
and unitarity then reads
\beq
\text{Im}(M(s,t))-\frac{1}{4\sqrt{-st}} |M(s,t)|^2 \le 0   \qquad \text{for} \qquad s=4m^2-t>4m^2 \,.
\eeq
This discussion will provide us with a simple example of numerics which work yet converge very slowly until we slightly improve our ansatz and thus completely solve this convergence issue. 
\begin{figure}
\begin{center}
\includegraphics[width=1 \linewidth]{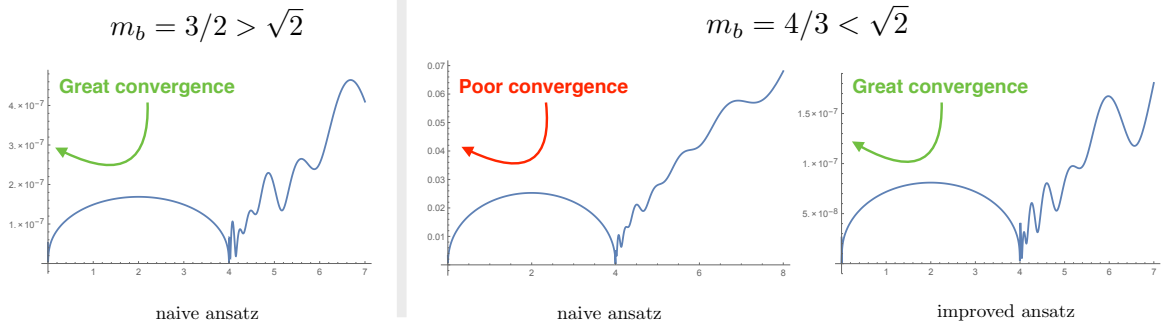}
\vspace{-8.5cm}
\caption{\label{convergence} Plot of $|(S_\text{num}-S_\text{analytic})/S_\text{analytic}|$, that is of the relative mismatch in the numerical solution in this two dimensional example where the analytic solution is available.
In all these plots we use $\Lambda=20$ and check unitarity in a small grid of $40$ points. With these parameters, \texttt{mathematica}'s built-in \texttt{FindMaximum} suffices and produces an outcome in about two or three seconds. We see on the left that for $m_b>\sqrt{2}m$ the agreement is spectacular with the most naive ansatz (\ref{naiveT}) while in the middle we see that the agreement is much worse (a few percent off) with the same ansatz when $m_b<\sqrt{2}m$. On the right we see that this is neatly fixed - leading again to a perfect convergence - by simply adopting an improved ansatz as in (\ref{improvedT}).  }
\end{center}
\end{figure}   

To be concrete we consider here the case where there is a single bound-state with mass $m_b$ whose coupling we maximize. The S-matrix with the largest coupling and such bound-state is given by \cite{paperII}
\beq
S_\text{max $g$} =\text{sign}(m_b-\sqrt{2} m)\times  \frac{\sqrt{s (s-4m^2)}+ \sqrt{m_b^2(4m^2-m_b^2)}}{\sqrt{s (s-4m^2)}- \sqrt{m_b^2(4m^2-m_b^2)}} 
\eeq
At high energies the S-matrix approaches $+1$ for $m_b>\sqrt{2}m$ and $-1$ for $m_b<\sqrt{2}m$ and this leads to a very different behavior when translated to the amplitude $M$. In particular, for a light bound state $m_b<\sqrt{2} m$ we see that  the amplitude $M$ in (\ref{STrelation}) must diverge at high energies so that the right hand side approaches $-2$. This is hard for an ansatz a la (\ref{repRho}) to achieve, that is it would require that the sum in 
\beq
M_\text{naive}(s,t) = -\frac{\hat g^2}{s-m_b^2}-\frac{\hat g^2}{t-m_b^2}+ \sum_{a,b=0}^{\Lambda} c_{ab}\, \rho_s^{a}\rho_t^{b} \la{naiveT}
\eeq
to develop a divergence as $s=4m^2-t\to \infty$ which corresponds to $\rho_s, \rho_t \to -1$. Such non-analytic behavior at the boundary of the unit disc \textit{can} be achieved \textit{but} a numerically sufficiently accurate approximation requires very large $\Lambda$.  

In this case there is however a very obvious improvement which is to simply allow for a divergence at large energies which is after all allowed by unitarity and write down instead an ansatz of the form 
\beq
M_\text{improved}(s,t) = -\frac{\hat g^2}{s-m_b^2}-\frac{\hat g^2}{t-m_b^2}+ \sum_{a,b=0}^{\Lambda-2} c_{ab}\, \rho_s^{a}\rho_t^{b} + \frac{\beta}{\sqrt{\rho_s+1}\sqrt{\rho_t+1}}+ \frac{\tilde\beta}{(\rho_s+1)(\rho_t+1)}
 \la{improvedT}
\eeq
This immediately allows for a more general high energy behavior and thus an extreme improvement in convergence as illustrated in figure \ref{convergence}.

The moral of this story seems to be that we better allow for flexible ansatze which can easily capture various analytic properties of scattering amplitudes if we want to achieve optimal convergence. In this simple two dimensional example, allowing for an ansatz with a more flexible high energy behavior led to a drastic
improvement in the numerics.

\bibliography{biblio}
\bibliographystyle{utphys}

\end{document}